\documentclass{article}

\usepackage{arxiv}          
\usepackage[utf8]{inputenc} 
\usepackage[T1]{fontenc}    
\usepackage{enumitem}
\usepackage[hidelinks]{hyperref}       
\usepackage{url}            
\usepackage{booktabs}       
\usepackage{amsfonts}       
\usepackage{nicefrac}       
\usepackage{microtype}      
\usepackage{amsmath}        
\usepackage{amsthm}         
\usepackage[normalem]{ulem} 
\usepackage{graphicx}
\usepackage{natbib}
\usepackage{placeins}
\usepackage{color}
\usepackage{textgreek}

\newtheorem{theorem}{Theorem}
\newtheorem{corollary}{Corollary}
\newtheorem{proposition}{Proposition}

\newtheorem{remark}{Remark}

\newtheoremstyle{assumption}%
  {}{}%
  {\itshape}{}%
  {\bfseries}{}%
  { }%
  {(\thmname{#1}\thmnumber{#2})}

\theoremstyle{assumption}
\newtheorem{assumption}{A}

\title{A Nonparametric Framework for Treatment Effect Modifier Discovery in High
  Dimensions}

\date{}

\author{
  Philippe Boileau \\
  Graduate Group in Biostatistics and\\
  Center for Computational Biology \\
  University of California, Berkeley\\
  \texttt{philippe\_boileau@berkeley.edu} \\
  \And
  Ning Leng \\
  Genentech Inc. \\
  \texttt{leng.ning@gene.com} \\
  \And
  Nima S. Hejazi \\
  Department of Biostatistics, \\
  T.H. Chan School of Public Health, \\
  Harvard University \\
  nhejazi@hsph.harvard.edu \\
  \And
  Mark van der Laan \\
  Division of Biostatistics, \\
  Department of Statistics, and\\
  Center for Computational Biology,\\
  University of California, Berkeley\\
  \texttt{laan@berkeley.edu} \\
  \And
  Sandrine Dudoit \\
  Department of Statistics, \\
  Division of Biostatistics, and \\
  Center for Computational Biology,\\
  University of California, Berkeley\\
  \texttt{sandrine@stat.berkeley.edu} \\
}

\renewcommand{\headeright}{Preprint}
\renewcommand{\undertitle}{Preprint}
\renewcommand{\shorttitle}{A Framework for Treatment Effect Modifier Discovery}

\begin{document}
\maketitle

\begin{abstract}
  Heterogeneous treatment effects are driven by treatment effect modifiers,
  pre-treatment covariates that modify the effect of a treatment on an outcome.
  Current approaches for uncovering these variables are limited to
  low-dimensional data, data with weakly correlated covariates, or data
  generated according to parametric processes. We resolve these issues by
  developing a framework for defining model-agnostic treatment effect modifier
  variable importance parameters applicable to high-dimensional data with
  arbitrary correlation structure, deriving one-step, estimating equation and
  targeted maximum likelihood estimators of these parameters, and establishing
  these estimators' asymptotic properties. This framework is showcased by
  defining variable importance parameters for data-generating processes with
  continuous, binary, and time-to-event outcomes with binary treatments, and
  deriving accompanying multiply-robust and asymptotically linear estimators.
  Simulation experiments demonstrate that these estimators' asymptotic
  guarantees are approximately achieved in realistic sample sizes for
  observational and randomized studies alike. This framework is applied to gene
  expression data collected for a clinical trial assessing the effect of a
  monoclonal antibody therapy on disease-free survival in breast cancer
  patients. Genes predicted to have the greatest potential for treatment effect
  modification have previously been linked to breast cancer. An open-source
  \texttt{R} package implementing this methodology, \texttt{unihtee}, is made
  available on GitHub at \url{https://github.com/insightsengineering/unihtee}.
\end{abstract}


\section{Introduction}\label{sec:intro}

The detection and quantification of heterogeneous treatment effects are central
to numerous areas of study in the medical and social sciences. Examples include
precision medicine, where practitioners seek patient subgroups exhibiting
differing benefits from a given therapy, and economics, where policymakers
assess the impact of government interventions on diverse population strata. This
heterogeneity is generally linked to treatment effect modifiers (TEM). TEMs are
pre-treatment covariates which, as their name suggests, modify the effect of a
treatment, alternatively referred to as an exposure, on the outcome. In
precision medicine, the response of patients with a shared disease to a common
therapy may be a function of, for example, sex-at-birth, age, genetic mutations,
and environmental exposures. Uncovering TEMs is therefore of great importance
when investigating or attempting to account for disparate effects of treatment
in a population.

Some parametric modeling techniques can accomplish just that in traditional
asymptotic settings under stringent conditions about the data-generating process
(DGP). When including treatment-covariate interaction terms in addition to main
effect terms in a linear model for a continuous outcome, TEMs are generally
defined as the features with non-zero interaction coefficients. Consistent
estimation and valid hypothesis testing of the TEMs are possible when the DGP
admits a linear relationship between the outcome, treatment, and covariates.
Generalized linear models (GLM) might be used for TEM discovery in more general
settings, such as when the outcome is binary or a non-negative integer. With
time-to-event outcomes, the Cox proportional hazards model with
treatment-covariate interactions might be used. If the posited functional
relationship does not correspond to reality, however, inference is invalid
\citep[see, for example,][]{hernan2010}.

Furthermore, the parameters corresponding to the aforementioned models, like the
odds ratio of a logistic regression model or the hazards ratio of a proportional
hazards model, depend on the other covariates included in the model. These
conditional parameters are said to be \textit{noncollapsible} \citep[for a
discussion and worked example, see][]{greenland1999}, in the sense that
marginalizing over the other covariates in the model may produce a marginal
parameter whose value differs from the marginal parameter directly obtained by
omitting these covariates (to see this, recall that for two random variables $X$
and $Y$ and an arbitrary function $g(\cdot)$, we generally find that
$\mathbb{E}[g(\mathbb{E}[Y \mid X])] \neq g(\mathbb{E}[\mathbb{E}[Y \mid X]]) = g(\mathbb{E}[Y])$).
Noncollapsible parameters lack a causal interpretation that unambiguously
relates them to marginal treatment effects.

More flexible approaches targeting the conditional average treatment effect
(CATE) may be employed to address these issues. In inferring the expected
difference in potential outcomes --- that is, the difference in outcomes that
could be computed if each observation's outcomes under treatment and control
were measured --- as a function of the covariates \citep{rubin1974}, CATE
estimators are uniquely suited for TEM discovery. The double-robust estimators
of \citet{zhao2018}, \citet{semenova2021}, and \citet{bahamyirou2022}, which
model the CATE using a linear model, permit valid statistical inference about
features' ability to modify the effect of treatment under less restrictive
assumptions about the DGP than traditional parametric methods. Others, like the
Super-Learner-based \citep{vdl2007} estimator of \citet{luedtke2016} or the
Random-Forests-inspired \citep{breiman2001} estimators of \citet{wager2018} and
\citet{cui2022}, rely on nonparametric supervised statistical learning
algorithms to identify potential TEMs under even fewer constraints on the DGP.

When the number of potential TEMs is commensurate with the number of
observations, or indeed much larger, the above parametric and CATE estimators'
capacity to reliably uncover TEMs diminishes. Estimation of the linear model
coefficients requires penalized regression methods like the LASSO
\citep{tibshirani1996,tian2014,chen2016}, rendering hypothesis testing of
treatment-covariate coefficients difficult. Practitioners might instead rely on
the asymptotic feature selection properties of the LASSO, but these hold only
under restrictive and unverifiable conditions on sparsity and covariate
correlation structures \citep{zhao2006}. Similar limitations plague the CATE
estimators relying on method-specific variable importance measures. In
particular, the causal forests of \citet{wager2018} and \citet{cui2022} can
assess the importance of variables, employing a permutation-based approach
analogous to those of traditional Random Forests. In high dimensions, however,
this metric can produce unreliable rankings of covariates' treatment
modification abilities: correlated features are likely to act as surrogates for
one another, leading to deflated importance scores \citep[Chap. 15]{esl2009}.

Instead of depending on algorithm-specific modeling strategies that treat TEM
discovery as a byproduct of conditional outcome or CATE estimation,
\citet{williamson2021framework, boileau2022}, and \citet{hines2022TEVIP}
recently proposed TEM variable importance parameters (TEM-VIP\footnote{Previous
  work has referred to TEM-VIPs as (treatment effect) variable importance
  measures (TE-)VIMs \citep[for example, ][]{williamson2021framework,
    hines2022TEVIP}, which we believe blurs the distinction between parameters
  and estimators and fails to emphasize that these measures of variable
  importance are well-defined parameters of a statistical model.}) that directly
assess the strength of covariates' capacity to modify the effect of treatment.
These algorithm-agnostic parameters, defined within nonparametric statistical
models and which may be augmented with causal interpretations, permit formal
statistical inference about TEMs.

Combining popular variable dropout procedures and previous work about the
variance of the conditional treatment effect estimator, \citet{levy2021},
\citet{williamson2021framework}, and \citet{hines2022TEVIP} proposed analogous
TEM-VIPs measuring individual or predefined sets of variables' influence on the
CATE variance. For instance, \citet{hines2022TEVIP} define the TEM-VIP of a set
of covariates as one minus the ratio of the variance of the CATE conditioning on
all but these covariates and the variance of the CATE conditioning on all
available covariates. The accompanying, nonparametric estimators are consistent
and asymptotically linear under nonrestrictive assumptions about the DGP.
However, these TEM-VIPs might produce misleading assessments of TEM impact when
the covariates are correlated: TEM-VIPs will generally possess values that do
not reflect covariates' capacity for treatment effect modification, like the
Random-Forests-based CATE variable importance measure of
\citet{wager2018,cui2022}. We expect this issue to be exacerbated in high
dimensions due to the increased chance of complex correlation structures.
Additionally, repeatedly omitting variables and estimating nuisance parameters
is computationally expensive --- and perhaps intractable --- when the number of
potential TEMs is large.

\citet{boileau2022} derived a marginal TEM-VIP expressly for high-dimensional
DGPs with continuous or binary outcomes. Assuming that the expected difference
in potential outcomes is linear in any given covariate when conditioning on said
covariate, the proposed TEM-VIP is the simple linear regression coefficient
obtained by regressing this difference on the potential TEM. \citet{boileau2022}
argue that this parameter provides a meaningful summary in all but pathological
DGPs: the larger it is, the larger the variables capacity for treatment effect
modification. Further, it does not suffer from the previously mentioned issues
associated with dropout- and permutation-based variable importance measures. A
nonparametric estimator of this TEM-VIP was proposed, and shown to be
double-robust and asymptotically linear under mild conditions on the DGP. A
simulation study demonstrated that these asymptotic properties were
approximately achieved in finite-sample, high-dimensional randomized control
trials.

This TEM-VIP is limited, however, to absolute effect modification for continuous
and binary responses. Expanding on this work and taking inspiration from
previous research on non- and semiparametric approaches \citep[for example,
][]{rosenblum2010, tchetgen2011, tuglus2011, chambaz2012, yadlowsky2019}, we
present a general framework for defining and performing inference about marginal
model-agnostic TEM-VIPs. Our approach is demonstrated through the creation of a
new absolute TEM-VIP for DGPs with right-censored time-to-event outcomes, and of
new relative TEM-VIPs for DGPs with continuous, binary, and right-censored
time-to-event outcomes. We derive one-step, estimating equation and targeted
maximum likelihood (TML) estimators based on these parameters' efficient
influence functions, study their asymptotic behavior, and investigate their
finite sample properties in simulation experiments. This general framework
equips practitioners with the tools to define bespoke TEM-VIPs, readily derive
nonparametric estimators, and establish sufficient conditions for which these
estimators permit reliable inference.

The remainder of the article is organized as follows:
Section~\ref{sec:cont-outcome} presents TEM-VIPs and related inference
procedures in data-generating processes with binary treatment variables and
continuous outcomes. The CATE-based TEM-VIPs of \citet{boileau2022} are
re-framed in terms of treatment effect modification discovery for continuous
outcomes in Section~\ref{subsec:cont-abs-vip}. Sufficient identifiability
conditions for the estimation of TEM-VIPs using observational data are also
presented, as are nonparametric estimators of this estimand. The asymptotic
properties of these estimators are then studied. A proposal for a relative
TEM-VIP follows in Section~\ref{subsec:cont-rel-vip}. Accompanying causal
identifiability conditions, nonparametric estimators, and sufficient conditions
for the desirable asymptotic behavior of these estimators are given.
Sections~\ref{sec:cont-outcome} and~\ref{sec:tte-outcome} introduce analogous
developments for data-generating processes with binary and time-to-event
outcomes, respectively. We discuss, in Section~\ref{sec:extending-framework},
the general procedure for defining model-agnostic TEM-VIPs, deriving
accompanying estimators, and studying their asymptotic characteristics in
nonparametric models. Simulation studies and a real data application are then
presented in Sections~\ref{sec:simulations} and \ref{sec:rct-data-application},
respectively, and we end with a discussion of our contributions in
Section~\ref{sec:discussion}. Proofs are relegated to the Appendix for clarity
of exposition.

\section{Continuous Outcomes}\label{sec:cont-outcome}

\subsection{Problem Setting}

Let there be $n$ independent and identically distributed (i.i.d.) random vectors
$\{X_i\}_{i=1}^n$, such that
$X_i = (W_i, A_i, Y_i^{(0)}, Y_i^{(1)}) \sim P_{X,0} \in \mathcal{M}_X$, where
$W_i$ is a set of $p$ covariates that are possibly treatment-outcome
confounders, $A_i$ is a binary variable indicating treatment assignment ($0$ for
control, $1$ for treatment), and $Y_i^{(1)}$ and $Y_i^{(0)}$ are continuous
potential outcomes \citep{rubin1974} that are assumed to be bounded such that
$Y_{i}^{(1)}, Y_{i}^{(0)} \in (0, 1)$ without loss of generality. The potential
outcomes $Y_i^{(1)}$ and $Y_i^{(0)}$ are the outcomes one would observe for the
$i^\text{th}$ observation had it been assigned to the treatment and control
conditions, respectively. Here, $p$ is of similar magnitude as, or larger than,
$n$. Finally, $\mathcal{M}_X$ is a nonparametric model of possible DGPs. We omit
the subscript $i$ where possible throughout the remainder of the text to ease
notational burden.

The true DGP, $P_{X,0}$, is generally unknown, and realizations of its random
vectors are typically unmeasurable, as only one potential outcome is observed.
Nevertheless, $P_{X,0}$ allows for the definition of causal parameters on which
statistical inference may subsequently be performed. An example of such a
parameter is the conditional average treatment effect (CATE):

\begin{equation*}
\mathbb{E}_{P_{X,0}}\left[Y^{(1)}-Y^{(0)} | W \right].
\end{equation*}

As discussed in Section~\ref{sec:intro}, however, the CATE poses a challenging
estimation problem --- even if somehow provided with the complete data generated
according to $P_{X,0}$ --- due to the dimension of $W$. Likewise, the recovery
of treatment effect modifiers using traditional variable importance techniques
based on CATE estimates, like penalized linear models or Random Forests
\citep{tian2014, chen2017, zhao2018, wager2018, ning2020, bahamyirou2022}, is
generally unreliable in high dimensions. We instead consider the causal TEM-VIP
proposed by \citet{boileau2022}.

\subsection{Absolute Treatment Effect Modification Variable Importance
  Parameter}\label{subsec:cont-abs-vip}

\subsubsection{Causal Parameter}

Indexing $W$ by $j=1, \ldots, p$, we place the following moment conditions on
$W$:
\begin{assumption}\label{ass:centered-confounders}
  Centered covariates: $\mathbb{E}_{P_{X,0}}[W_{j}] = 0$, without loss of generality.
\end{assumption}
\begin{assumption}\label{ass:non-zero-variance}
  Non-zero variance: $\mathbb{E}_{P_{X,0}}[W_{j}^{2}] > 0$, for $j=1,\ldots, p$.
\end{assumption}
An absolute TEM-VIP of the $j$\textsuperscript{th} covariate is then defined as
a mapping
\begin{equation}\label{eq:cont-abs-vip}
  \Psi_j^{F}(P_{X,0}) \equiv
  \frac{\text{Cov}_{P_{X,0}}[Y^{(1)}-Y^{(0)}, W_{j}]}
  {\mathbb{V}_{P_{X,0}}[W_{j}]}
  = \frac{\mathbb{E}_{P_{X,0}}\left[
  \left(Y^{(1)}-Y^{(0)}\right)W_j\right]}
  {\mathbb{E}_{P_{X,0}}\left[W_j^2\right]} \;.
\end{equation}
Letting $\bar{Q}_{P_{X,0}}(a,W) \equiv \mathbb{E}_{P_{X,0}}[Y^{(a)}|W]$, it is
straightforward to show that
\begin{equation*}
  \Psi_j^{F}(P_{X,0}) = \frac{\mathbb{E}_{P_{X,0}}\left[
  \left(\bar{Q}_{P_{X,0}}(1,W)-\bar{Q}_{P_{X,0}}(0, W)\right)W_j\right]}
  {\mathbb{E}_{P_{X,0}}\left[W_j^2\right]} \;.
\end{equation*}
The estimand is then given by
$\Psi^{F}: \mathcal{M}_{X} \rightarrow \mathbb{R}^{p}$,
$\Psi^F(P_{X,0}) = (\Psi_1^F(P_{X,0}), \ldots, \Psi_p^F(P_{X,0}))$.

A\ref{ass:centered-confounders} lightens notation throughout the manuscript and
simplifies inference about $\Psi^{F}_{j}$. To the see the latter, consider that
$\text{Cov}_{P_{X,0}}[Y^{(1)}-Y^{(0)}, W_{j}] = \mathbb{E}_{P_{X,0}}[W_{j}(Y^{(1)}-Y^{(0)})] - \mathbb{E}_{P_{X,0}}[W_{j}]\mathbb{E}_{P_{X,0}}[Y^{(1)}-Y^{(0)}]$.
If $\mathbb{E}[W_{j}] \neq 0$, the average treatment effect is a nuisance
parameter. When A\ref{ass:centered-confounders} is not satisfied, however,
covariates can be demeaned using their sample mean. This similarly simplifies
inference about $\Psi^{F}_{j}$ and has no bearing on the asymptotic results of
its estimators presented later in the manuscript.

A\ref{ass:non-zero-variance} is easily satisfied in practice by filtering
variables exhibiting no variability. Note too that pre-treatment covariates with
zero variance cannot possibly modify the effect of the treatment.

Assuming the expectation of $\bar{Q}_{P_{X,0}}(1,W)-\bar{Q}_{P_{X,0}}(0, W)$
conditional on any given $W_j$ is linear in $W_j$, $\Psi^F(P_{X,0})$ is the
vector of simple linear regression coefficients generated by regressing the
differences in expected potential outcomes against the individual elements of
$W$. That is, let $f(W) = \bar{Q}_{P_{X,0}}(1,W)-\bar{Q}_{P_{X,0}}(0, W)$, and
assume that $\mathbb{E}_{P_{X,0}}[f(W)|W_{j}] = \beta_{j}W_{j}$. Then, for
$j=1,\ldots,p$,
\begin{equation*}
    \Psi_{j}^{F}(P_{X,0})
    = \frac{\mathbb{E}_{P_{X,0}}\left[f(W)W_{j}\right]}
      {\mathbb{E}_{P_{X,0}}\left[W_{j}^{2}\right]}
    = \frac{\mathbb{E}_{P_{X,0}}\left[
      \mathbb{E}_{P_{X,0}}\left[f(W)W_{j}|W_{j}\right]\right]}
      {\mathbb{E}_{P_{X,0}}\left[W_{j}^{2}\right]}
    = \frac{\mathbb{E}_{P_{X,0}}\left[\beta_{j}W_{j}^{2}\right]}
      {\mathbb{E}_{P_{X,0}}\left[W_{j}^{2}\right]}
    =  \beta_{j} \;.
\end{equation*}
When the relationship between $f(W)$ and the $W_{j}$'s is nonlinear, as is
  generally the case, this parameter can be interpreted as a linear model projection
  in the sense of the nonparametric marginal structural models of
  \citet{neugebauer2007}:
  \begin{equation*}
    \Psi_{j}^{F}(P_{X,0}) =
    \text{arg min}_{\beta_{j} \in \mathbb{R}} \mathbb{E}_{P_{X,0}}\left[
      \bigg(\left(\bar{Q}_{P_{X,0}}(1,W)-\bar{Q}_{P_{X,0}}(0, W)\right) -
        \left(\alpha + \beta_{j} W_{j}\right)\bigg)^{2}
    \right] \; .
  \end{equation*}
  Here, $\alpha$ is the average treatment effect. Thus, $\Psi^{F}(P_{X,0})$ can
  be viewed as assessing the correlation between the difference in potential
  outcomes and each potential TEM, re-normalized to be on the same scale as
  $Y$.

Regardless, a positive (negative) $\Psi_{j}^{F}(P_{X,0})$ suggests that
observations with larger absolute realizations of $W_{j}$ experience more
positive (negative) treatment effects.This parameter is not without its
limitations, however. As mentioned in \citet{boileau2022},
$\Psi_{j}^{F}(P_{X,0})$ may fail to detect treatment effect modifiers that have
pathological nonlinear relationships with the difference in potential outcomes.
If such relationships are suspected, TEM-VIPs of derived covariates, defined as
demeaned, nonlinear transformations of the initial covariate set, could be
considered.

Though TEM-VIPs provide a continuous measure of the strength of the treatment
effect modifications, some applications may call for the dichotomization of
covariates into TEMs and non-TEMs based on $\Psi^{F}(P_{X,0})$. By default, we
classify the $j$\textsuperscript{th} covariate $W_{j}$ as a TEM if
$\lvert \Psi^{F}_{j}(P_{X,0})\rvert > 0$ and note that, in some settings, it may
make sense to impose a non-zero threshold for this classification. Of course,
using a common non-zero threshold is only reasonable when the covariates have
comparable units. If this is not the case, then covariates might be scaled using
their sample variances prior to performing inference about $\Psi^{F}(P_{X,0})$.
As with the de-meaning of the covariates, this scaling will not influence the
asymptotic results presented later in the manuscript.

\subsubsection{Identifiability Through Observed-Data Parameter}

The full data
$\{X_i\}_{i=1}^n = \{(W_{i}, A_{i}, Y_{i}^{(0)}, Y_{i}^{(1)})\}_{i=1}^{n}$ are
generally censored through the treatment assignment mechanism. We instead have
access to $n$ i.i.d. random variables $O = (W, A, Y) \sim P_0 \in \mathcal{M}$.
The statistical model, $\mathcal{M}$, is fully determined by $\mathcal{M}_{X}$:
for each $P_{X} \in \mathcal{M}_{X}$, there exists a unique $P \in \mathcal{M}$,
where $W$ and $A$ are defined as in the full-data DGP and $Y$, the observed
outcome variable, is given by $AY^{(1)} + (1-A)Y^{(0)}$ such that $Y \in (0, 1)$
without loss of generality. This relationship between $Y$, $Y^{(1)}$, and
$Y^{(0)}$ is often referred to as the consistency
assumption~\citep{hernan2023causal}. Here, $P_0$ is the unknown DGP of the
observed data. Throughout the remainder of the text, we denote the empirical
distribution by $P_n$, the expected conditional outcome
$\mathbb{E}_{P_0}[Y \mid A,W]$ by $\bar{Q}_0(A,W)$, and the propensity score
$\mathbb{P}_{P_0}[A = 1 \mid W]$ by $g_0(W)$. $\bar{Q}_0(A,W)$ and $g_0(W)$ are
written as $\bar{Q}_0$ and $g_0$ where possible for notational convenience.

Now, the challenge lies in establishing an equivalence between a parameter of
the DGP for the observed data $O$ and $\Psi^F(P_{X,0})$. \citet{boileau2022}
provided sufficient identifiability conditions for just that. We present them
here, along with a formal statement:
\begin{assumption}\label{ass:no-unmeasured-confounding}
  No unmeasured confounding: $Y^{(a)} \perp A|W$, for $a \in \{0, 1\}$.
\end{assumption}
\begin{assumption}\label{ass:positivity}
  Positivity: there exists some constant $\epsilon > 0$ such that
  $\mathbb{P}_{P_0}[\epsilon < g_0(W) < 1-\epsilon] = 1$.
\end{assumption}
\begin{theorem}\label{thm:id-cond-abs-risk}
  Assuming that A\ref{ass:centered-confounders}, A\ref{ass:non-zero-variance},
  A\ref{ass:no-unmeasured-confounding}, and A\ref{ass:positivity} hold, we find
  that
  \begin{equation}\label{eq:stat-param-cont-abs-risk}
    \Psi_j(P_0)
    \equiv \frac{\mathbb{E}_{P_0}\left[
        \left(\bar{Q}_0(1,W)-\bar{Q}_0(0, W)\right)W_j\right]}
        {\mathbb{E}_{P_0}\left[W_j^2\right]}
    = \Psi_j^F(P_{X,0}),
  \end{equation}
  for $j=1,\ldots, p$. The parameter
  $\Psi: \mathcal{M} \rightarrow \mathbb{R}^p$ defined as
  $\Psi(P_0) = (\Psi_1(P_0), \ldots, \Psi_p(P_0))$ is therefore equal to the
  full-data estimand $\Psi^F(P_{X,0})$.
\end{theorem}

The two latest assumptions are ingrained in the causal inference literature.
Assumption A\ref{ass:no-unmeasured-confounding} ensures that treatment
assignment is regarded as if performed in a randomized experiment. It is more
easily satisfied by considering many pre-treatment covariates as potential
confounders, especially when the complete set of confounders is unknown.
Assumption A\ref{ass:positivity} requires all observations to have a non-zero
probability of receiving either treatment condition, guaranteeing that
$\bar{Q}_{0}(1, W)$ and $\bar{Q}_{0}(0, W)$ are equal to
$\bar{Q}_{P_{X,0}}(1,W)$ and $\bar{Q}_{P_{X,0}}(0,W)$, respectively.

\begin{remark}\label{remark:known-confounders}
  Assumptions A\ref{ass:no-unmeasured-confounding} and A\ref{ass:positivity} may
  be relaxed when the confounding variables are known \textit{a priori}.
  Consider that $W$ is the union of the --- possibly overlapping --- sets of
  candidate TEMs and confounders. Denoting the random vector of confounders by
  $V$, the assumption of no unmeasured confounding is satisfied when
  $Y^{(a)} \perp A | V$ for $a \in \{0, 1\}$. Further,
  $\mathbb{P}_{P_{0}}[A=1|W] = \mathbb{P}_{P_{0}}[A=1|V]$ such that the
  positivity assumption may be re-written as
  $\mathbb{P}_{P_{0}}[\epsilon < \mathbb{P}_{P_{0}}[A=1|V] < 1-\epsilon] = 1$
  for $\epsilon > 0$.
\end{remark}

\subsubsection{Inference}

Having established an identifiable parameter, we detail procedures for
performing inference about it. We first, however, briefly review the basics of
nonparametric asymptotic theory.

\paragraph{Preliminaries}

Consider a degenerate distribution $\tilde{P}$ that places all support of its
random observations $\tilde{O}$ on $\tilde{o}$. Further assume that $\tilde{o}$
is contained in the support of $P \in \mathcal{M}$ and define a two-component
mixture model $P_{\epsilon} = \epsilon \tilde{P} + (1-\epsilon) P$. Appealing to
Riesz's representation theorem \citep{fisher2021,hines2022}, the efficient
influence function of a parameter $\Theta(P)$ is defined through the following
Gateaux --- that is, functional --- derivative:
\begin{equation}\label{eq:generic-eif}
  \frac{d\Theta(P_{\epsilon})}{d\epsilon} \bigg|_{\epsilon = 0}
    = \lim_{\epsilon \rightarrow 0}
    \frac{\Theta(P_{\epsilon}) - \Theta(P)}{\epsilon}
    = \int \theta(o)(d\tilde{P}(o) - dP(o))
    = \int D(o, P)d\tilde{P}(o)
    = D(\tilde{o}, P)\;.
\end{equation}
Here, $\theta$ is defined such that $\mathbb{E}_{P}[\theta(O)] = \Theta(P)$, and
$D(o, P) = \theta(o) - \int \theta(o) dP(o) = \theta(o) - \Theta(P)$. $D(O,P)$
therefore generalizes the concept of directional derivatives to functionals,
measuring $\Theta$'s sensitivity to perturbations of $P$. Intuitively, then,
$\mathbb{E}_{P}[D(O,P)] = 0$ for $O \sim P$. When the variance of $D(O,P)$ is
bounded under all $P \in \mathcal{M}$, we say that the Gateaux derivative is
well-defined and that $\Theta$ is pathwise differentiable.

Now, similar to how asymptotic approximations of mean-based parameters are
studied through Taylor expansions, the asymptotic behavior of the plug-in
estimator $\Theta(\hat{P}_{n})$, a functional, is studied by way of a von Mises
expansion \citep{mises1947asymptotic, bickel1993efficient, vdl2003unified,
  hines2022}. Here, $\hat{P}_{n}$ is a plug-in estimator of $P_{0}$ that is made
up of elements of the $P_{n}$ and, possibly, nuisance parameter estimators. This
functional equivalent to the Taylor expansion is defined in terms of the
efficient influence function (EIF):

\begin{equation}\label{eq:von-mises}
    \sqrt{n}\left(\Theta(\hat{P}_n)-\Theta(P_0)\right)
    = \sqrt{n}\mathbb{E}_{P_n}\left[D(O,P_0)\right]
      - \sqrt{n}\mathbb{E}_{P_n}\left[D(O,\hat{P}_n)\right]
    + \sqrt{n}\left(\mathbb{E}_{P_{n}}-\mathbb{E}_{P_{0}}\right)
      \left(D(O,\hat{P}_n)-D(O,P_0)\right)
      - \sqrt{n}R(P_0,\hat{P}_n) \;.
\end{equation}
$R(P_0, \hat{P}_n)$ is a second-order remainder term in $n$. Since
$\mathbb{E}_{P_0}[D(O, P_0)] = 0$, the first term of
Equation~\eqref{eq:von-mises} converges to a Gaussian random variable with mean
zero and variance equal to $\mathbb{E}_{P_0}[D(O,P_{0})^2]$ by the central limit
theorem. The second term is a bias term that generally does not vanish
asymptotically. The third term of the von Mises expansion can generally be shown
to converge to zero in probability under sufficient empirical process
conditions. Alternatively, sample-splitting procedures, often referred to as
\textit{cross-fitting}, can be used to relax these conditions, as suggested by
\citet{pfanzagl1985,klaasen1987,zheng2011cross,cherno2017}. The remainder term
can generally be shown to converge to zero in probability under convergence rate
assumptions about nuisance parameter estimators.

Nonparametric estimators, like the one-step \citep{pfanzagl1985,
  bickel1993efficient}, estimating equation \citep{vdl2003unified, cherno2017},
and targeted maximum likelihood (TML) estimators \citep{vdl2006targeted,
  vdl2011targeted, vdl2018targeted}, correct the asymptotic bias term. They are
constructed from the efficient influence function.
\begin{enumerate}
  \item [One-Step] This estimator is derived by subtracting the
        asymptotic bias term from the plug-in estimator:
        $\Theta^{(\text{OS})}(\hat{P}_{n}) \equiv
        \Theta(\hat{P}_{n}) + \mathbb{E}_{P_{n}}[D(O, \hat{P}_{n})]$.
  \item [Estimating Equation] $\Theta^{(\text{EE})}(\hat{P}_n)$ is the solution to
        the following estimating equation: $0 = \mathbb{E}_{P_{n}}[D(O, \hat{P}_{n})]$.
  \item [TML] $\Theta^{(\text{TML})}(\hat{P}_{n})$ is obtained by tilting
        $\hat{P}_{n}$ to generate a $P_{n}^{\star}$ such that
        $\mathbb{E}_{P_{n}}\left[D(O, P_{n}^{\star})\right] \approx 0$. There
        are many ways to achieve this. Examples are provided later in the text,
        as well as in \citet{vdl2006targeted, vdl2011targeted, vdl2018targeted}.
        The estimator is then defined as
        $\Theta^{(\text{TML})}(\hat{P}_{n}) \equiv \Theta(P_{n}^{\star})$, the
        plug-in estimator using $P_{n}^{\star}$. Unlike one-step and estimating
        equation estimators, TML estimators constrain estimates to the parameter
        space.
\end{enumerate}

Provided the required conditions ensuring the third and fourth terms of
Equation~\eqref{eq:von-mises} converge in probability to zero are met, these
estimators are asymptotically linear and efficient. That is, they are
asymptotically normally distributed with mean $\Theta(P_{0})$ and variance
$\mathbb{E}_{P_0}[D(O,P_{0})^2/n]$ and have the smallest asymptotic variance
among regular and asymptotically linear (RAL) estimators in a nonparametric
model \citep{bickel1993efficient, tsiatis2006, vdl2011targeted}.

Inference about $\Theta(P_0)$ can then be based on the asymptotically normal
distribution of its one-step, estimating equation, and TML estimators. In
particular, the $\alpha$-level Wald-type confidence interval for $\Theta(P_0)$
can be constructed identically for each of the three estimators,
$\Theta^{(\cdot)}(\hat{P}_{n})$, as follows:
\begin{equation*}
  \Theta^{(\cdot)}(\hat{P}_n) \pm
  z_{1-\alpha/2}\sqrt{\frac{\mathbb{E}_{P_0}[D(O,P_0)^2]}{n}},
\end{equation*}
where $z_{1-\alpha/2}$ is the $(1-\alpha/2)^\text{th}$ quantile of the standard
Normal distribution. Of course, $\mathbb{E}_{P_0}[D(O,P_0)^2]$ is generally
unknown; an estimator, $\mathbb{E}_{P_n}[D(O, \hat{P}_n)^{2}]$, is used instead.
When there are many tests to perform in small-to-moderate sample sizes, the
empirical Bayes approach to variance estimation proposed by \citet{hejazi2022}
might also be employed for improved Type I error rate control.

\paragraph{Efficient Influence Function}

We return to our study of the TEM-VIP $\Psi(P_{0})$, defined in Equation
\eqref{eq:stat-param-cont-abs-risk}. The efficient influence function of
$\Psi_j(P)$ for $P\in\mathcal{M}$ was previously derived by \citet{boileau2022}.
We restate it here for convenience.

\begin{proposition}\label{prop:eif-cont-abs-vip}
  Assume A\ref{ass:centered-confounders} and A\ref{ass:non-zero-variance}.
  Define $\Psi_j(P)$ as in Equation~\eqref{eq:stat-param-cont-abs-risk} for some
  $P \in \mathcal{M}$. The efficient influence function at $P \in \mathcal{M}$
  of this parameter is given by
  \begin{equation}\label{eq:eif-cont-abs-risk}
      D_j(O, P) \equiv \frac{W_j}{\mathbb{E}_P\left[W_j^2\right]}
        \left(\frac{2A-1}{Ag(W) + (1-A)(1-g(W))}
        \left(Y - \bar{Q}(A,W)\right)
        + \bar{Q}(1, W)
        - \bar{Q}(0, W) - \Psi_j(P)W_j\right) \;.
  \end{equation}
\end{proposition}

\paragraph{Estimators}

We now present nonparametric estimators of the TEM-VIP of Equation
\eqref{eq:stat-param-cont-abs-risk}.

{\em One-step and estimating equation estimators.} The one-step estimator of
$\Psi_j(P_0)$, for $j=1,\ldots, p$, is identical to the estimating equation
estimator $\Psi_j^{(\text{ee})}(\hat{P}_n)$ \citep{boileau2022}. Let $\bar{Q}_n$
and $g_n$ be, respectively, estimators of $\bar{Q}_0$ and $g_0$ trained on $P_n$
and included in $\hat{P}_{n}$. Then
\begin{equation*}
  \Psi^{(\text{OS})}_j(\hat{P}_n) = \Psi^{(\text{EE})}_{j}(\hat{P}_{n})
  \equiv \frac{1}{\sum_{i=1}^n W_{ij}^2}
  \sum_{i=1}^n W_{ij} \bigg(
  \frac{2A_{i}-1}{A_{i}g_{n}(W_{i}) + (1-A_{i})(1-g_{n}(W_{i}))}
  \left(Y_i - \bar{Q}_n(A_i,W_i)\right)
  + \bar{Q}_n(1, W_i) - \bar{Q}_{n}(0, W_{i}) \bigg) \; .
\end{equation*}

{\em Targeted maximum likelihood estimator.} The TML estimator's derivation is
slightly more involved. Recall that $Y \in (0, 1)$. Define the negative
log-likelihood loss function for $\bar{Q}$ as
\begin{equation*}
  L(O;\bar{Q}) \equiv -\log \left\{\bar{Q}(A,W)^{Y}
    \left(1-\bar{Q}(A,W)\right)^{(1-Y)}\right\} \; ,
\end{equation*}
and a parametric working submodel for $\bar{Q}$ as
\begin{equation*}
  \bar{Q}_{j}(\epsilon)(A,W) \equiv \text{logit}^{-1}\left\{
    \text{logit}\; \bar{Q}(A,W) + \epsilon H_{j}(A,W)
  \right\},
\end{equation*}
where
\begin{equation*}
  H_{j}(A,W) \equiv \frac{W_{j}}{\mathbb{E}_{P}[W_{j}^{2}]}
    \frac{2A-1}{Ag(W) + (1-A)(1-g(W))} \; .
\end{equation*}
Now, denoting an initial estimator of $\bar{Q}_0$ trained on $P_{n}$ by
$\bar{Q}_n^0$, we update $\bar{Q}_n^0$ by computing $\epsilon_{n,j}^1$ such
that
\begin{equation*}
  \epsilon_{n,j}^1 = \text{arg min}_{\epsilon} \; \mathbb{E}_{P_{n}}\left[
    L(O; \bar{Q}_{n,j}^0(\epsilon))\right],
\end{equation*}
where, though not immediately clear in the notation, $\bar{Q}_{n,j}^0(\epsilon)$
depends directly on $\bar{Q}_n^0$ and indirectly (through $H_{j}(A,W)$) on $g_n$
and $\sum_{i} W_{ij}^2/n$, an estimator of $\mathbb{E}_{P}[W_{j}^{2}]$. A tilted
conditional outcome estimator is then computed as
$\bar{Q}_{n,j}^1 \equiv \bar{Q}_{n,j}^0(\epsilon_{n,j}^1)$. The solution to the
above equation, $\epsilon_{n,j}^{1}$, is the maximum likelihood estimator (MLE)
of a univariate logistic regression's slope coefficient obtained by regressing
$Y$ on $H_{n,j}(A,W)$ while taking $\bar{Q}_{n}^{0}(A,W)$ as an offset. Here,
$H_{n,j}$ is the empirical version of $H_{j}$, using $g_{n}$ and
$\sum_{i} W_{ij}^{2}/n$ in place of $g$ and $\mathbb{E}_{P}[W_{j}^{2}]$,
respectively.

We define $P_{n}^{\star}$ as the tilted $\hat{P}_{n}$, where $\bar{Q}_{n}^{0}$
is replaced by $\bar{Q}_{n,j}^1$. Exploiting a classical result of logistic
regression in parametric statistical models \citep{vdl2006targeted}, it follows
that $\mathbb{E}_{P_{n}}[D(O, P_{n}^{\star})] \approx 0$. The TML estimator of
the $j^\text{th}$ TEM-VIP is therefore given by
$\Psi_j^\text{(\text{TML})}(\hat{P}_n) \equiv \Psi_j(P_n^{\star})$.

We highlight that this estimator is appropriate even when the outcome is not
restricted to $(0, 1)$. It suffices to shift each of the observed outcomes
$Y_{i}$, $i=1, \ldots, n$, by $-\min_{i}\{Y_{i}\}$ and to scale them by
$\max_{i}\{Y_{i}\} - \min_{i}\{Y_{i}\}$ prior to computing the TML estimate, and
then rescaling the TML estimate by the same quantities. Note too that other loss
functions might be used to tilt $\hat{P}_{n}$, like the squared error loss. See
\citet{gruber2010} for a comparison and discussion.

\paragraph{Asymptotic Behavior}

Next, we study the asymptotic behavior of these absolute TEM-VIP estimators.
Note that the asymptotic distributions of $\Psi^{\text{(OS)}}(\hat{P}_n)$,
$\Psi^{\text{(EE)}}(\hat{P}_n)$, and $\Psi^{\text{(TML)}}(\hat{P}_n)$ are
identical through their dependence on $D_j(O,P_0)$, $j=1,\ldots,p$. In
particular, they are double-robust, meaning they are consistent even when one of
the nuisance parameters is inconsistently estimated.

\begin{assumption}\label{ass:consistency-cond-outcome}
  Conditional outcome estimator consistency:
  \begin{equation*}
    \lVert \bar{Q}_n(A,W) - \bar{Q}_0(A,W)\rVert_2^{2} =
    \int  (\bar{Q}_n(a,w) - \bar{Q}_0(a,w))^2 dP_0(a,w) = o_P(1).
\end{equation*}
\end{assumption}
\begin{assumption}\label{ass:consistency-prop-score}
  Propensity score estimator consistency:
  \begin{equation*}
    \lVert g_n(W) - g_0(W) \rVert_2^2 = \int  (g_n(w) - g_0(w))^2 dP_0(w) = o_P(1).
  \end{equation*}
\end{assumption}

\begin{proposition}\label{prop:dr-cont-abs-risk}
  Under A\ref{ass:centered-confounders} and A\ref{ass:non-zero-variance}, and
  either A\ref{ass:consistency-cond-outcome} or
  A\ref{ass:consistency-prop-score},
  $\mathbb{E}_{P_{0}}[D_j(O, \hat{P}_{n})|\hat{P}_{n}] = o_{P}(1)$ for
  $j=1,\ldots,p$. That is,
  $\Psi^{(\text{OS})}(\hat{P}_{n}) \overset{P}{\rightarrow} \Psi(P_{0})$, and
  the same is true of the estimating equation and TML estimators.
\end{proposition}

Further, these estimators' sampling distribution can be specified under the
following assumptions:
\begin{assumption}\label{ass:donsker}
  Donsker conditions: There exists a $P_0$-Donsker class\footnote{A class
    $\mathcal{F}$ with bounded supremum norm is P-Donsker if $\mathcal{F}$ is
    pre-Gaussian and the empirical process $\mathbb{G}(\mathcal{F})$ converges
    weakly under $L_{\infty}(P)$ to the Gaussian process
    $\mathbb{G}_P(\mathcal{F})$ in $n$. Here,
    $\mathbb{G}(\mathcal{F}) = \{\mathbb{G}(f), f \in \mathcal{F}\}$
    \citep{bickel1993efficient, vdl2011targeted, bickel2015mathstats}.}
  $\mathbb{G}_0$ such that
  $\mathbb{P}_{P_{0}}[D_j(O, \hat{P}_n) \in \mathbb{G}_0] \rightarrow 1$ and
  $\mathbb{E}_{P_{0}}[(D_j(O,\hat{P}_n)-D_j(O,P_0))^2|\hat{P}_{n}] = o_P(1)$ for
  each $j$.
\end{assumption}
\begin{assumption}\label{ass:double-rate-robustness}
  Shared rate convergence:
  $\lVert \bar{Q}_n(A,W) - \bar{Q}_0(A,W)\rVert_2\;
  \lVert g_n(W) - g_0(W)\rVert_2 = o_P(n^{-1/2})$.
\end{assumption}
\begin{assumption}\label{ass:bounded-confounders}
  Bounded covariates: There exists $C \in
  \mathbb{R}_{+}$, such that $\lvert W_{j} \rvert \leq C$ for $j=1,\ldots,p$.
\end{assumption}
\begin{theorem}\label{thm:asymp-cont-abs-risk-ests}
  Assuming A\ref{ass:centered-confounders}, A\ref{ass:non-zero-variance},
  A\ref{ass:donsker}, A\ref{ass:double-rate-robustness}, and
  A\ref{ass:bounded-confounders},
  $\sqrt{n}(\Psi_{j}^{(\text{OS})}(\hat{P}_{n}) - \Psi_j(P_0)) = (1/\sqrt{n}) \sum_i D_{j}(O_{i},P_{0}) + o_{P}(1)$
  for $j = 1, \ldots, p$. This implies that
  $\sqrt{n}(\Psi_{j}^{(\text{OS})}(\hat{P}_{n}) - \Psi_j(P_0)) \overset{D}{\rightarrow} N(0, \mathbb{E}_{P_0}[D_j(O,P_0)^{2}])$.
  This result is true of the estimating equation and TML estimators, too.
\end{theorem}

A\ref{ass:donsker} is a generally unverifiable assumption guaranteeing that the
third term on the right-hand side of Equation~\eqref{eq:von-mises} converges to
zero in probability. However, it is equivalent to placing weak regularity
conditions on $\bar{Q}$ and $g$ that are more transparent. If these nuisance
parameters are c\`{a}dl\`{a}g (\textit{continue \`{a} droite, limite \`{a}
  gauche}: right continuous, left limits) \citep{neuhaus1971} and have finite
supremum and (sectional) variation norms \citep{gill1995inefficient}, for
example, then A\ref{ass:donsker} is met \citep[see][for a discussion of these
conditions]{vdl2016, vdl2017}. Alternatively, nonparametric estimators may be
extended to perform a sample-splitting procedure such that this requirement is
replaced by even milder conditions \citep{bickel1982,schick1986, klaasen1987,
  zheng2011cross, cherno2017}, like the convergence of $\bar{Q}_n$ or $g_n$ to
fixed functionals that are not necessarily equal to $\bar{Q}_0$ and $g_0$,
respectively \citep{zheng2011cross}. \citet{boileau2022} derive such a
cross-fitted estimator based on the one-step approach.

A\ref{ass:double-rate-robustness} is satisfied when the nuisance parameter
estimators jointly converge at the standard semiparametric rate of $n^{-1/2}$.
This so-called ``shared rate convergence'' condition also allows for one of the
nuisance parameters to converge more slowly if the other estimator converges
more quickly. When $p$ is small relative to $n$,
A\ref{ass:double-rate-robustness} is typically satisfied by estimating
$\bar{Q}_0$ and $g_0$ with flexible machine learning methods that place few, if
any, assumptions on the functional form of these parameters. One such approach
is the Super Learner framework \citep{vdl2007}. In high-dimensional
observational settings, however, this convergence property is only met by
appealing to smoothness and sparsity assumptions about the nuisance parameters.
Examples of these conditions for Random Forests and deep neural networks are
outlined in \citet{wager2018} and \citet{farrell2021}, respectively. Of note,
A\ref{ass:double-rate-robustness} is satisfied regardless of $p$'s size relative
to $n$ when either of the nuisance parameters are known, as is the case with
$g_0$ in randomized controlled trials (RCT). This follows from inspection of the
second-order remainder term of Equation~\eqref{eq:von-mises}, which equals to
zero in this scenario.

We note that A\ref{ass:donsker} and A\ref{ass:double-rate-robustness} are
satisfied when estimating the nuisance parameters with the Highly Adaptive LASSO
under the condition that these parameters are c\`{a}dl\`{a}g and have bounded
sectional variation norm \citep{vdl2017, bibaut2019}. Current implementations of
this estimator, available in the \texttt{hal9001}
R package~\citep{hejazi2020hal9001, coyle-cran-hal9001}, are currently too
computationally demanding for use with the high-dimensional DGPs considered,
however.

The final assumption, A\ref{ass:bounded-confounders}, is a sufficient, technical
condition required to bound the second-order remainder of
Equation~\eqref{eq:von-mises}. While it may appear stringent, and is, for
example, not satisfied by covariates generated according to a multivariate
Gaussian distribution, we believe that it is generally applicable. Many --- if
not most --- of the random variables studied in the biological, physical, and
social sciences are bounded by their very nature or by limitations of
measurement instruments. We demonstrate
Theorem~\ref{thm:asymp-cont-abs-risk-ests}'s practical robustness to
A\ref{ass:bounded-confounders} in the simulation experiments of
Section~\ref{sec:simulations} by generating covariates with Gaussian
distributions.

We remark that while Theorem~\ref{thm:asymp-cont-abs-risk-ests} states that
$\Psi^{(\text{OS})}(\hat{P}_n)$ and $\Psi^{(\text{TML})}(\hat{P}_n)$ --- as well
as their cross-fitted counterparts --- are asymptotically identical, noticeable
differences in behavior are possible in finite samples. This is explored in the
simulation study of Section~\ref{sec:simulations}.

\subsection{Relative Treatment Effect Modification Variable Importance Parameter}\label{subsec:cont-rel-vip}

\subsubsection{Causal Parameter}

While the TEM-VIP of Equation~\eqref{eq:cont-abs-vip} is a generally informative
assessment of treatment effect modification, situations may arise where a
relative TEM-VIP is of greater interest. Examples include DGPs with non-negative
outcome variables or when investigating the effect modification of relative
treatment effects.

Consider a parameter based on the ratio of conditional
expected potential outcomes:
\begin{equation*}
  \frac{\mathbb{E}_{P_{X,0}}\left[Y^{(1)}\big|W\right]}
    {\mathbb{E}_{P_{X,0}}\left[Y^{(0)}\big|W\right]} \\
  = \frac{\bar{Q}_{P_{X,0}}(1,W)}{\bar{Q}_{P_{X,0}}(0,W)} \; .
\end{equation*}
The full-data model, $\mathcal{M}_X$, is identical to the one presented in
Section~\ref{subsec:cont-abs-vip}, save that
$Y^{(0)}, Y^{(1)} \in \mathbb{R}_+$. As with the CATE, estimating the
conditional parameter above is challenging in high dimensions, making TEM
discovery unreliable. Again assuming A\ref{ass:centered-confounders} and
A\ref{ass:non-zero-variance}, we instead propose a TEM-VIP inspired by a GLM of
the outcome with a log link function:
\begin{equation}\label{eq:cont-rel-vip}
  \Gamma_{j}^F(P_{X,0}) \equiv
  \frac{\mathbb{E}_{P_{X,0}}\left[
      \left(\log\bar{Q}_{P_{X,0}}(1,W)
        -\log\bar{Q}_{P_{X,0}}(0, W)\right)W_j\right]}
  {\mathbb{E}_{P_{X,0}}\left[W_j^2\right]} \; .
\end{equation}
Then $\Gamma^{F}: \mathcal{M}_{X} \rightarrow \mathbb{R}^{p}$,
$\Gamma^{F}(P_{X,0}) = (\Gamma^{F}_{1}(P_{X,0}), \ldots, \Gamma^{F}_{p}(P_{X,0}))$
is the target of inference.

Assuming that the expectation of
$\log\bar{Q}_{P_{X,0}}(1,W)-\log\bar{Q}_{P_{X,0}}(0, W)$ conditional on any
given $W_{j}$ is linear in $W_{j}$,
$\Gamma^F(P_{X,0}) = (\Gamma_{1}^F(P_{X,0}), \ldots, \Gamma_{p}^F(P_{X,0}))$ is
the vector of simple linear regression coefficients produced by regressing the
log-ratio of expected conditional potential outcomes against individual
covariates. As with $\Psi^{F}(P_{X,0})$, $\Gamma^{F}(P_{X,0})$ remains an
informative estimand under violations of this linearity assumption in all but
pathological scenarios, and can be viewed as assessing the correlation between
the log-ratio of potential outcomes and each covariate. As in the absolute
TEM-VIP case, $W_{j}$ is said to be a TEM under this relative VIP if
$\lvert \Gamma_{j}^{F}(P_{X,0})\rvert > 0$.

\subsubsection{Identifiability Through Observed-Data Parameter}

Relating $\Gamma^{F}(P_{X,0})$ to some parameter of $P_{0}$ follows directly from
the result of Theorem~\ref{thm:id-cond-abs-risk}.
\begin{corollary}\label{cor:id-cond-rel-risk}
  Under the conditions outlined in Theorem~\ref{thm:id-cond-abs-risk},
  \begin{equation}\label{eq:stat-param-cont-rel-risk}
    \Gamma_j(P_0)
    \equiv \frac{\mathbb{E}_{P_0}\left[
        \left(\log\bar{Q}_0(1,W)-\log\bar{Q}_0(0, W)\right)W_j\right]}
        {\mathbb{E}_{P_0}\left[W_j^2\right]}
     = \Gamma_j^F(P_{X,0}),
  \end{equation}
  for $j=1,\ldots, p$. The observed-data parameter
  $\Gamma: \mathcal{M} \rightarrow \mathbb{R}^p$ defined as
  $\Gamma(P_0) = (\Gamma_1(P_0), \ldots, \Gamma_p(P_0))$ is therefore equal to
  the full-data estimand $\Gamma^F(P_{X,0})$.
\end{corollary}

\subsubsection{Inference}

\paragraph{Efficient Influence Function}

To lighten notation, $D_{j}(O,P)$ is recycled to represent the efficient
influence function of $\Gamma_{j}(P)$ and all other parameters throughout the
remainder of the manuscript.

\begin{proposition}\label{prop:eif-cont-rel-vip}
  Assume A\ref{ass:centered-confounders} and A\ref{ass:non-zero-variance}, and
  define $\Gamma_j(P)$ as in Equation~\eqref{eq:stat-param-cont-rel-risk} for
  $P\in\mathcal{M}$. The efficient influence function of this parameter is
  \begin{equation}\label{eq:eif-cont-rel-vip}
      D_j(O,P)
       \equiv \frac{W_j}{\mathbb{E}_P\left[W_j^2\right]}\Bigg(
        \frac{2A-1}{Ag(W) + (1-A)(1-g(W))}
        \frac{Y-\bar{Q}(A,W)}{\bar{Q}(A,W)}
        + \log\frac{\bar{Q}(1, W)}{\bar{Q}(0,W)} - \Gamma_j(P)W_j\Bigg) \; .
  \end{equation}
\end{proposition}

\paragraph{Estimators}

Nonparametric estimators of $\Gamma(P_{0})$ are given next.

{\em One-step and estimating equation estimators.} From the von Mises expansion
of Equation~\eqref{eq:von-mises}, we find that the one-step TEM-VIP estimator
for the $j^\text{th}$ potential TEM is given by
\begin{equation*}
  \Gamma_j^{(\text{OS})}(\hat{P}_n)
  \equiv \frac{1}{\sum_{i=1}^n W_{ij}^2}
  \sum_{i=1}^n W_{ij} \left(
    \frac{2A_{i}-1}{A_{i}g_n(W_{i}) + (1-A_{i})(1-g_{n}(W_{i}))}
    \frac{Y_i - \bar{Q}_n(A_i,W_i)}{\bar{Q}_n(A_i,W_i)}
    + \log\frac{\bar{Q}_n(1, W_i)}{\bar{Q}_n(0, W_i)}\right) \;.
\end{equation*}
As with the absolute TEM-VIP, the estimating equation estimator of
$\Gamma(P_0)$, $\Gamma_{j}^{(\text{EE})}(\hat{P}_n)$, is identical to
$\Gamma_{j}^{(\text{OS})}(\hat{P}_n)$.

{\em Targeted maximum likelihood estimator.} The TML estimator of
$\Gamma_{j}(P_{0})$, $\Gamma_{j}^{\text{(TML)}}(\hat{P}_{n})$, is computed using
a targeting strategy that is almost identical to that of
$\Psi_{j}^{\text{(TML)}}(\hat{P}_{n})$. The only departure from the previously
presented procedure occurs in the definition of $H_{j}(A,W)$. For the relative
TEM-VIP, we let
\begin{equation*}
  H_j(A,W) \equiv \frac{W_{j}}{\mathbb{E}_{P}[W_{j}^{2}]\bar{Q}(A,W)}
  \frac{2A-1}{Ag(W) + (1-A)(1-g(W))} \; .
\end{equation*}
The calculation of the tilted estimator $\bar{Q}_{n,j}^{1}$ is otherwise
unchanged. It then follows that
$\Gamma_{j}^{\text{(TML)}}(\hat{P}_{n})\equiv\Gamma_{j}(P_{n}^{\star})$, where
we again stress through notation that $\Gamma_{j}^{\text{(TML)}}(\hat{P}_{n})$
is a plug-in estimator relying on the tilted distribution $P_{n}^{\star}$.
$P_{n}^{\star}$ is identical to $\hat{P}_{n}$ save that $\bar{Q}_{n,j}^{1}$ is
used in place of $\bar{Q}_{n}$.

\paragraph{Asymptotic Behavior}

As before, we begin the study of $\Gamma^{(\text{OS})}(\hat{P}_n)$,
$\Gamma^{(\text{EE})}(\hat{P}_n)$, and $\Gamma^{\text{(TML)}}(\hat{P}_n)$'s
identical asymptotic behavior with sufficient conditions for consistency.

\begin{proposition}\label{prop:cons-cont-rel-risk}
  If A\ref{ass:centered-confounders}, A\ref{ass:non-zero-variance}, and
  A\ref{ass:consistency-cond-outcome} are satisfied,
  $\mathbb{E}_{P_{0}}[D_j(O,\hat{P}_{n})|\hat{P}_{n}] = o_{P}(1)$ for
  $j=1,\ldots,p$. That is,
  $\Gamma^{(\text{OS})}(\hat{P}_{n}) \overset{P}{\rightarrow} \Gamma(P_{0})$.
  This result holds for the estimating equation and TML estimators as well.
 \end{proposition}

We contrast Propositions~\ref{prop:dr-cont-abs-risk} and
\ref{prop:cons-cont-rel-risk}: Unlike $\Psi^{(\text{OS})}$ and
$\Psi^{(\text{TML})}$, $\Gamma^{(\text{OS})}$ and $\Gamma^{(\text{TML})}$ are
not doubly robust. Consistent estimation of $\bar{Q}_{0}$ is required and,
barring random positivity violations, estimation of $g_{0}$ has no impact.

Next, the asymptotic linearity of these estimators is established.

\begin{assumption}\label{ass:rate-consistency-cond-outcome}
  Convergence rate of conditional outcome estimator:
  $\lVert \bar{Q}_n(A,W) - \bar{Q}_0(A,W)\rVert_2 = o_P(n^{-1/4})$.
\end{assumption}
\begin{theorem}\label{thm:asymp-cont-rel-risk-ests}
  Under A\ref{ass:centered-confounders}, A\ref{ass:non-zero-variance},
  A\ref{ass:donsker}, A\ref{ass:double-rate-robustness},
  A\ref{ass:bounded-confounders}, and A\ref{ass:rate-consistency-cond-outcome},
  $\sqrt{n}(\Gamma_{j}^{(\text{OS})}(\hat{P}_{n}) - \Gamma_j(P_0)) = (1/\sqrt{n}) \sum_i D_j(O_{i},P_0) + o_{P}(1)$.
  Again, this result applies to the estimating equation and TML estimators, and
  implies that
  $\sqrt{n}(\Gamma_{j}^{(\text{OS})}(\hat{P}_{n}) - \Gamma_j(P_0)) \overset{D}{\rightarrow} N(0, \mathbb{E}_{P_0}[D_j(O,P_0)^{2}])$.
\end{theorem}

The conditions required for the asymptotic linearity of
$\Gamma^{(\text{OS})}(\hat{P}_{n})$ and $\Gamma^{(\text{TML})}(\hat{P}_{n})$ are
largely similar to those of $\Psi^{(\text{OS})}(\hat{P}_{n})$ and
$\Psi^{(\text{TML})}(\hat{P}_{n})$. The sole difference is that candidate
estimators of the conditional expected outcome must converge at a rate no slower
than $o_{P}(n^{-1/4})$. The propensity score estimator, however, may converge at
a slower rate so long as A\ref{ass:double-rate-robustness} is satisfied. While
this distinction has little impact in observational study settings, the same
cannot be said in RCTs. Knowing $g_{0}$ does not guarantee the asymptotic
linearity of $\Gamma^{(\text{OS})}(\hat{P}_n)$ and
$\Gamma^{(\text{TML})}(\hat{P}_n)$; an accurate estimator of $\bar{Q}_0$ is
essential. Future investigations may assess whether this parameter's efficient
influence function possesses double-robust properties in alternative models.
Indeed, we suspect this to be the case in more restrictive semiparametric models
imposing parametric assumptions on $Y \mid A,W$.

\begin{remark}
Consider the setting identical to that described in this section, save that the
outcome, $Y$, is a binary random variable. Noting that $\bar{Q}(A, W)
= \mathbb{P}[Y = 1 \mid A, W]$, it follows that all results of
Section~\ref{sec:cont-outcome} apply to these DGPs. That is, the absolute and
relative TEM-VIPs, as well as their respective asymptotically linear estimators,
can just as readily be used to detect treatment effect modifiers when the
outcome is binary.
\end{remark}

\section{Right-Censored Time-to-Event Outcomes}\label{sec:tte-outcome}

Returning to the motivating example of the introduction, the discovery of
treatment effect modifiers is essential to precision medicine: they delineate
patient subgroups, allowing for tailored care. They can also provide mechanistic
insight on experimental therapies and improve the success rate of clinical
trials. However, the data generated and collected in many therapeutic areas,
like oncology, are characterized censored time-to-event outcomes like time to
death or disease recurrence. The TEM-VIPs presented thus far are not readily
applicable to this setting.

\subsection{Problem Setting}\label{subsec:tte-prob-setting}

Consider $n$ i.i.d.~random vectors $\{X_{i}\}_{i=1}^{n}$, where
$X = (W, A, C^{(0)}, C^{(1)}, T^{(0)}, T^{(1)}) \sim P_{X,0} \in \mathcal{M}_{X}$.
We again define $\mathcal{M}_{X}$ as a nonparametric statistical model of
possible full-data DGPs and denote the true DGP by $P_{X,0}$. As before, $W$ and
$A$ are, respectively, the vector of pre-treatment covariates and the binary
treatment indicator. Here, $C^{(a)}$ and $T^{(a)}$ correspond, respectively, to
the (discrete or continuous) censoring and event times, from which we define the
right-censored time-to-event $\tilde{T}^{(a)} = \min \{T^{(a)}, C^{(a)}\}$ and
the censoring indicator $\Delta^{(a)} = I(T^{(a)} > C^{(a)})$, under condition
$a \in \{0,1\}$.

Causal parameters of interest in this setting often build upon the conditional
survival function
$S_{P_{X,0}}(t|a, W) \equiv \mathbb{P}_{P_{X,0}}[T^{(a)} > t | W]$,
$a \in \{0,1\}$. Consider the CATE of the survival probability at time $t$:
\begin{equation*}
  \mathbb{E}_{P_{X,0}}\left[S_{P_{X,0}}(t|1, W) - S_{P_{X,0}}(t|0, W)| W \right] \;.
\end{equation*}
The difference in conditional restricted mean survival times (RMST)
\citep{chen2001,royston2011} for time $t$ might be a meaningful target causal
parameter too:
\begin{equation*}
  \mathbb{E}_{P_{X,0}}\left[
    \min \{T^{(1)}, t\} - \min \{T^{(0)}, t\} \Big| W
  \right]
  = \mathbb{E}_{P_{X,0}}\left[
    \int_{0}^{t} \left\{S_{P_{X,0}}(u|1, W) - S_{P_{X,0}}(u|0, W)\right\} du \Big|W
  \right] \;.
\end{equation*}
A derivation of the above equality is found in \citet{diaz2019}.

As with the CATE in DGPs with continuous and binary outcomes, however, the
recovery of treatment effect modifiers from these parameters is unreliable in
high dimensions. We suggest using the TEM-VIPs described in the subsequent
subsections instead.

\subsection{Absolute Treatment Effect Modification Variable Importance
  Parameter}\label{subsec:tte-abs-vip}

\subsubsection{Causal Parameter}

Under A\ref{ass:centered-confounders} and A\ref{ass:non-zero-variance}, the
following measure of absolute treatment effect modification for
time-to-event outcomes can be used:
\begin{equation}\label{eq:tte-abs-vip}
  \Psi_{j}^{F}(P_{X,0}; t) \equiv \frac{
    \mathbb{E}_{P_{X,0}}\left[
      W_{j} \int_{0}^{t}\left\{S_{P_{X,0}}(u|1,W)-S_{P_{X,0}}(u|0,W)\right\}du\right]
  }{\mathbb{E}_{P_{X,0}}\left[W_{j}^{2}\right]} \;.
\end{equation}
The estimand is then given by
$\Psi^{F}:\mathcal{M}_{X} \times \mathbb{R}_{+} \rightarrow \mathbb{R}^{p}$,
$\Psi^{F}(P_{X,0}; t) = (\Psi_{1}^{F}(P_{X,o};t), \ldots, \Psi^{F}_{p}(P_{X,0};t))$.
We reuse ``$\Psi$'' to emphasize that this is an absolute effect parameter,
noting that $t \mapsto \Psi^{F}_{j}(P; t)$ defines a curve.

Similar to the continuous outcome scenario, $\Psi_{j}^{F}(P_{X,0}; t)$ captures
the correlation of the difference in conditional RMSTs and the
$j$\textsuperscript{th} covariate, standardized to be on the outcome's scale.
$\Psi^{F}(P_{X,0};t) = (\Psi_{1}^{F}(P_{X,0};t),\ldots,\Psi_{p}^{F}(P_{X,0};t))$
therefore generally identifies the pre-treatment covariates responsible for the
largest differences in expected truncated survival times.

\subsubsection{Identifiability Through Observed-Data Parameter}

As before, the full data $\{X_{i}\}_{i=1}^{n}$ are typically not observable.
Define $T = AT^{(1)}+(1-A)T^{(0)}$ and $C = AC^{(1)}+(1-A)C^{(0)}$. We instead
have access to $\{O_{i}\}_{i=1}^{n}$, a set of $n$ random variables
$O = (W, A, \tilde{T}, \Delta) \sim P_{0} \in \mathcal{M}$, where $W$ and $A$
are defined as in the full-data model,
$\tilde{T} = \min\{T,C\} = A\min\{T^{(1)}, C^{(1)}\} + (1-A)\min\{T^{(0)}, C^{(0)}\}$
is the right-censored time-to-event, and $\Delta = I(T > C)$ is the censoring
indicator. Again, $P_0$ is the true unknown DGP for the observed data $O$ and is
fully specified by $P_{X,0}$, and $\mathcal{M}$ is the model of possible
observed-data DGPs. Further, let $S_{0}(t|A,W) \equiv \mathbb{P}_{P_0}[T>t|A,W]$
and $\mathbb{P}_{P_{0}}[C>t|A,W] \equiv c_{0}(t|A,W)$ represent the observed
conditional survival and censoring functions, respectively.

Sufficient identifiability conditions relating $\Psi^{F}(P_{X,0};t)$ to a
parameter of the observed-data DGP are provided next.

\begin{assumption}\label{ass:no-unmeasured-confounding-t-a}
  No unmeasured exposure-time-to-event confounding: $T^{(a)} \perp A|W$, for
  $a \in \{0, 1\}$. Unclear how to interpret $A$ in this and the next
  assumption.
\end{assumption}
\begin{assumption}\label{ass:no-unmeasured-confounding-t-c}
  No unmeasured time-to-event-censoring confounding:
  $T^{(a)} \perp C^{(a)}|A,W$, for $a \in \{0, 1\}$.
\end{assumption}
\begin{assumption}\label{ass:censoring-positivity}
  Censoring mechanism positivity: There exists some $\epsilon > 0$ such that
  $\mathbb{P}_{P_{0}}[c_{0}(u| A, W) < 1-\epsilon] = 1$ for all $u \in
  (0, t)$.
\end{assumption}

\begin{theorem}\label{thm:id-tte-abs-vip}
  Assuming A\ref{ass:centered-confounders}, A\ref{ass:non-zero-variance},
  A\ref{ass:positivity}, A\ref{ass:no-unmeasured-confounding-t-a},
  A\ref{ass:no-unmeasured-confounding-t-c}, and A\ref{ass:censoring-positivity}
  hold, we find that
  \begin{equation}\label{eq:id-tte-abs-vip}
      \Psi_{j}(P_{0}; t)
      \equiv \frac{
        \mathbb{E}_{P_0}\left[
        W_{j} \int_{0}^{t}\left\{S_{0}(u|1, W)-S_{0}(u|0,W)\right\}du\right]
        }{\mathbb{E}_{P_0}\left[W_{j}^{2}\right]}
      = \Psi_{j}^{F}(P_{X,0}; t) \; ,
  \end{equation}
  for $j=1,\ldots,p$. The observed-data parameter
  $\Psi : \mathcal{M}\times \mathbb{R}_{+} \rightarrow \mathbb{R}^{p}$,
  $\Psi(P_{0}; t) = (\Psi_{1}(P_{0}; t), \ldots, \Psi_{p}(P_{0}; t))$ is equal
  to $\Psi^{F}(P_{X,0};t)$.
\end{theorem}

Beyond the condition that the covariates be centered and have non-zero variance,
the assumptions required by Theorem~\ref{thm:id-tte-abs-vip} are standard in the
causal inference literature for time-to-event parameters \citep[see, for
example,][]{moore2011, benkeser2019, diaz2019}.
A\ref{ass:no-unmeasured-confounding-t-a} ensures that the treatment assignment
mechanism can be viewed as random, conditional on the covariates.
A\ref{ass:no-unmeasured-confounding-t-c} requires that survival and censoring
times are independent given treatment and covariates. Finally,
A\ref{ass:censoring-positivity} specifies that every random unit has a positive
probability of being observed at every time up to and including $t$. Like the
identifiability results of Theorem~\ref{thm:id-cond-abs-risk} and
Remark~\ref{remark:known-confounders}, assumptions
A\ref{ass:no-unmeasured-confounding-t-a},
A\ref{ass:no-unmeasured-confounding-t-c}, and A\ref{ass:censoring-positivity}
may be modified when the sets of confounders are known.

\subsubsection{Inference}

\paragraph{Efficient Influence Function}

The efficient influence function of the estimand in
Equation~\eqref{eq:id-tte-abs-vip} is provided below.

\begin{proposition}\label{prop:eif-tte-abs-vip}
  Define $\Psi_{j}(P;t)$ as in Equation~\eqref{eq:id-tte-abs-vip} for some
  $P \in \mathcal{M}$ and assume A\ref{ass:centered-confounders} and
  A\ref{ass:non-zero-variance}. The uncentered efficient influence function of
  $S(t|a,W)$ is given by
  \begin{equation*}
    d(O,P;t, a) \equiv \frac{I(A=a) S(t|a,W)}{(Ag(W) + (1-A)(1-g(W)))}
    \int_{0}^{t} \frac{I(\tilde{T} \geq u)}{c(u_{-}|a,W) S(u|a,W)}
    \left(I(T=u)-\lambda(u|a,W)\right) du + S(t|a,W) \;,
  \end{equation*}
  where $\lambda(u|A,W)$ is the conditional survival hazard at time $u$ and
  $u_{-}$ denotes the left-hand limit of $u$ \citep{moore2011}. By the
  functional delta method, the efficient influence function of $\Psi_{j}(P; t)$
  is
  \begin{equation}\label{eq:eif-tte-abs-vip}
      D_{j}(O,P;t)
      \equiv \frac{W_{j}}{\mathbb{E}_{P}\left[W_{j}^{2}\right]}
        \left(\int_{0}^{t} d(O, P; u, 1) - d(O, P; u, 0) \; du
        - \Psi_{j}(P; t)W_{j}
        \right) \; .
  \end{equation}
\end{proposition}

\paragraph{Estimators}

In practice, for numerical reasons, the integrals in the estimators presented
next are approximated by weighted sums.

{\em One-step and estimating equation estimators.} It follows immediately from
Proposition~\ref{prop:eif-tte-abs-vip} that the one-step and estimating equation
estimators of $\Psi_{j}(P_{0}; t)$ are defined as
\begin{equation*}
  \Psi_{j}^{(\text{OS})}(\hat{P}_n; t) = \Psi_{j}^{(\text{EE})}(\hat{P}_n; t) \equiv
    \frac{1}{\sum_{i=1}^{n} W_{ij}^{2}} \left(
      \sum_{i=1}^{n} W_{ij} \int_{0}^{t}
         d(O_{i},\hat{P}_n;u, 1)- d(O_{i},\hat{P}_n;u, 0) du
    \right) \;.
\end{equation*}

{\em Targeted maximum likelihood estimator.} Let the log-likelihood loss of
$\lambda(u|A,W)$ be given by
\begin{equation*}
  L(O;\lambda, u) = -\log \left\{
    \lambda(u|A,W)^{I(T=u)}(1-\lambda(u|A,W))^{1-I(T=u)}
  \right\} \;.
\end{equation*}
Define the parametric working submodel for $\lambda(u|A,W)$ as
\begin{equation*}
  \lambda(\epsilon)(u|A, W) = \text{logit}^{-1}\{\text{logit} \lambda(u|A,W) +
    \epsilon H_{j}(u|A,W)\} \;,
\end{equation*}
where
\begin{equation*}
  H_{j}(u|A,W) \equiv \frac{W_{j}(2A-1)S(u|A,W)}
  {\left(Ag(W)+(1-A)(1-g(W))\right) \mathbb{E}_{P}\left[W_{j}^{2}\right]}
    \int_{0}^{u}
    \frac{1}{c(v_{-}|A,W)S(v|A,W)}
    dv \;.
\end{equation*}
Denoting the initial estimator of $\lambda_{0}(u|A,W)$ by
$\lambda_{n}^{0}(u|A,W)$, we update $\lambda_{n}^{0}(u|A,W)$ by computing
$\epsilon_{n}^{1}$, where
\begin{equation*}
  \epsilon^{1}_{n,j} = \text{arg min}_{\epsilon}
  \mathbb{E}_{P_{n}}\left[L(O; \lambda_{n}^{0}(\epsilon), u)\right] \;.
\end{equation*}
This empirical expectation is minimized using the MLE of the univariate logistic
regression of the event indicators $(1-\Delta) I(\tilde{T} = v)$ on
$H_{n,j}(u|A,W)$, for time $v$ ranging from $0$ to $u$ and with the initial
hazard estimates as an offset. $H_{n,j}$ is the empirical counterpart of
$H_{j}$, using $S_{n}, g_{n}, c_{n}$, and $\sum_i W_{ij}^{2}/n$ in place of
$S, g, c$, and $\mathbb{E}_{P}[W_{j}^{2}]$, respectively. The longitudinal
structure of the data need not be considered \citep{moore2011}; the repeated
measures are treated as independent when estimating $\epsilon_{n,j}^{1}$
\citep{moore2011}. $\lambda_{n,j}^{1}(u|A,W)$ is then defined as
$\lambda_{n}(\epsilon_{n,j}^{1})(u|A,W)$. Setting
$\lambda_{n,j}^{0}(u|A,W) \leftarrow \lambda_{n,j}^{1}(u|A,W)$, this procedure
is repeated until $\epsilon^{1}_{n,j} \approx 0$.

This procedure for tilting the conditional hazard at time $u$ is performed at
each observed time point between $0$ and $t$. These tilted hazards replace their
initial counterparts in $\hat{P}_n$ to form the tilted empirical distribution
$P_{n}^{\star}$. Noting that
$S_{n,j}(t|A,W) = \Pi_{u=1}^{t}(1-\lambda_{n,j}^{1}(u|A,W))$, it follows that
the TML estimator of $\Psi_{j}(P_{0};t)$ is given by
$\Psi_{j}^{(\text{TML})}(\hat{P}_n;t)\equiv\Psi_{j}(P_{n}^{\star}, t)$.

\paragraph{Asymptotic Behavior} We now consider the asymptotic properties of
these estimators.

\begin{assumption}\label{ass:cond-surv-consistency}
  Conditional survival estimator consistency:
  \begin{equation*}
    \lVert S_{n}(u|A,W)-S_{0}(u|A,W)\rVert_{2}^{2} =
    \int (S_n(u|a,w) - S_0(u|a,w))^2 dP_0(a,w) =
    o_{P}(1)
  \end{equation*}
  for all $u \in [0, t]$.
\end{assumption}
\begin{assumption}\label{ass:g-consistency}
  Conditional propensity score estimator and censoring estimator consistency:
  $\lVert g_{n}(W) - g_{0}(W)\rVert_{2}^{2} = o_{P}(1)$ and
  \begin{equation*}
    \lVert c_{n}(u|A,W)-c_{0}(u|A,W)\rVert_{2}^{2} =
    \int (c_n(u|a,w) - c_0(u|a,w))^2 dP_0(a,w) =
    o_{P}(1)
  \end{equation*}
  for all $u \in [0, t]$.
\end{assumption}

\begin{proposition}\label{prop:dr-tte-abs-vip-est}
  $\Psi^{(\text{OS})}(\hat{P}_n; t) \overset{P}{\rightarrow} \Psi(P_{0}; t)$ when
  A\ref{ass:centered-confounders}, A\ref{ass:non-zero-variance}, and either
  A\ref{ass:cond-surv-consistency} or A\ref{ass:g-consistency} are satisfied.
  This result also applies to $\Psi^{(\text{EE})}(\hat{P}_n; t)$ and
  $\Psi^{(\text{TML})}(\hat{P}_n; t)$.
\end{proposition}

\begin{assumption}\label{ass:consistency-treatment-survival}
  Shared convergence rate:
  $\lVert g_{n}(W)-g_{0}(W)\rVert_{2}\lVert S_{n}(u|A,W)-S_{0}(u|A,W)\rVert_{2} = o_{P}(n^{-1/2})$
  for all $u \in [0, t]$.
\end{assumption}
\begin{assumption}\label{ass:consistency-censoring}
  Convergence rate of conditional censoring estimator:
  $\lVert c_{n}(u|A,W)-c_{0}(u|A,W)\rVert_{2} = o_{P}(n^{-1/4})$ for all
  $u \in [0, t]$.
\end{assumption}

\begin{theorem}\label{thm:asymp-tte-abs-risks-ests}
  Assuming that A\ref{ass:centered-confounders}, A\ref{ass:non-zero-variance},
  A\ref{ass:donsker}, A\ref{ass:bounded-confounders},
  A\ref{ass:consistency-treatment-survival}, and
  A\ref{ass:consistency-censoring} are met,
  $\sqrt{n}(\Psi_{j}^{(\text{OS})}(\hat{P}_n;t) - \Psi_{j}(P_{0}; t)) = (1/\sqrt{n}) \sum_i D_{j}(O_{i}, P_{0}; t) + o_{P}(1)$.
  The same is true for the estimating equation and TML estimators. Again, this
  implies that
  $\sqrt{n}(\Psi_{j}^{(\text{OS})}(\hat{P}_n;t) - \Psi_{j}(P_{0}; t))\overset{D}{\rightarrow} N(0, \mathbb{E}_{P_{0}}[D_{j}(O, P_{0}; t)^{2}])$.
\end{theorem}

Proposition~\ref{prop:dr-tte-abs-vip-est} states that consistent estimation of
the TEM-VIPs is possible if either the conditional survival function is
consistently estimated or if the treatment assignment mechanism and the
censoring mechanism are consistently estimated. This implies that, in an RCT,
consistent estimates of $\Psi(P_{0}; t)$ only require that the censoring
mechanism be consistently estimated. When there is no censoring or censoring is
known to be independent of covariates, consistency is guaranteed when
$c_{0}(t|A,W)=c_{0}(t|A)$ is estimated with the Kaplan-Meier estimator.

Enforcing more stringent conditions on the DGP and the nuisance parameter
estimators results in Theorem~\ref{thm:asymp-tte-abs-risks-ests}. That is,
requiring that A\ref{ass:donsker} is satisfied --- or, alternatively, that
nuisance parameters are estimated via cross-fitting --- and that the nuisance
parameters estimators are consistent at the rates given in
A\ref{ass:consistency-treatment-survival} and A\ref{ass:consistency-censoring}
ensures asymptotically normal estimators that are centered around the true
parameter value. When the treatment assignment mechanism is known, as in an RCT,
then the only necessary consistency rate condition is that of the censoring
mechanism. Valid inference is therefore possible even when the conditional
survival function is misspecified.

\subsection{Relative Treatment Effect Modification Variable Importance
  Parameter}\label{subsec:tte-rel-vip}

\subsubsection{Causal Parameter}

As mentioned in Section~\ref{subsec:cont-abs-vip}, a relative TEM-VIP may be of
greater relevance than an absolute TEM-VIP in some contexts. In particular, when
treatment effect modification is assessed in terms of conditional probabilities,
as is done in this time-to-event setting, a relative measure may be more
sensitive. We propose a causal parameter analogous to that of
Equation~\eqref{eq:cont-rel-vip}:
\begin{equation}\label{eq:tte-rel-vip}
  \Gamma_{j}^{F}(P_{X,0}; t) \equiv \frac{
    \mathbb{E}_{P_{X,0}}\left[\left(\log S_{P_{X,0}}(t|1, W)-
        \log S_{P_{X,0}}(t|0,W)\right)W_{j}\right]
  }{\mathbb{E}_{P_{X,0}}\left[W_{j}^{2}\right]} \;.
\end{equation}
Again, we assume A\ref{ass:centered-confounders} and
A\ref{ass:non-zero-variance}. Then
$\Gamma^{F}: \mathcal{M}_{X} \times \mathbb{R}_{+} \rightarrow \mathbb{R}$,
$\Gamma^{F}(P_{X,0};t)=(\Gamma_{1}^{F}(P_{X,0};t),\ldots, \Gamma_{p}^{F}(P_{X,0};t))$
can be interpreted in a similar fashion to the relative TEM-VIP of the
continuous outcome DGP. As in Section~\ref{subsec:tte-abs-vip}, ``$\Gamma$'' is
reused to stress that this is a relative parameter.

\subsubsection{Identifiability Through Observed-Data Parameter}

The causal TEM-VIP $\Gamma^{F}(P_{X,0}; t)$ is identifiable in the observed data
under the conditions outlined in Theorem~\ref{thm:id-tte-abs-vip}. This follows
immediately given that $S_{0}(t|A,W) = S_{P_{X,0}}(t|A,W)$.

\begin{corollary}\label{cor:id-tte-rel-vip}
  Under the assumptions of Theorem~\ref{thm:id-tte-abs-vip}, it follows that
  \begin{equation}\label{eq:id-tte-rel-vip}
      \Gamma_{j}(P_0; t)
      \equiv \frac{
        \mathbb{E}_{P_0}\left[\left(\log S_{0}(t|1,W)-
        \log S_{0}(t|0,W)\right)W_{j}\right]
        }{\mathbb{E}_{P_0}\left[W_{j}^{2}\right]}
      = \Gamma_{j}^{F}(P_{X,0}; t)
  \end{equation}
  such that $\Gamma: \mathcal{M}\times\mathbb{R}_{+}\rightarrow\mathbb{R}^p$,
  $\Gamma(P_{0}; t) = (\Gamma_{1}(P_0; t), \ldots, \Gamma_{p}(P_0; t)) = \Gamma^{F}(P_{X,0},t)$.
\end{corollary}

\subsubsection{Inference}

\paragraph{Efficient Influence Function}

The efficient influence function of the observed-data parameter presented in
Equation~\eqref{eq:id-tte-rel-vip} is given next.

\begin{proposition}\label{prop:eif-tte-rel-vip}
  Assuming A\ref{ass:centered-confounders} and A\ref{ass:non-zero-variance}, the
  efficient influence function of $\Gamma(P; t)$ is
  \begin{equation}\label{eq:eif-tte-rel-vip}
    \begin{split}
      D_{j}(O,P;t)
      & \equiv \frac{W_{j}} {\mathbb{E}_{P}\left[W_{j}^{2}\right]}
        \left(\frac{2A-1}{Ag(W) + (1-A)(1-g(W))}
        \int_{0}^{t}
        \frac{I(\tilde{T} \geq u)}
        {c(u_{-}|A,W) S(u|A,W)}
        \left(I(T=u)-\lambda(u|A,W)\right) du \right.\\
      & \qquad\qquad\qquad\qquad + \log \frac{S(t|1,W)}{S(t|0,W)}
        - \Gamma(P;t)W_{j}\Bigg) \;.
    \end{split}
  \end{equation}
\end{proposition}

\paragraph{Estimators}

{\em One-step and estimating equation estimators.}
$\Gamma^{(\text{OS})}(\hat{P}_n;t)$ and $\Gamma^{(\text{EE})}(\hat{P}_n;t)$, are then
given by
\begin{equation*}
  \begin{split}
    \Gamma^{(\text{OS})}(\hat{P}_n;t) = \Gamma^{(\text{EE})}(\hat{P}_n;t)
    & \equiv \frac{1}{\sum_{i=1}^{n} W_{ij}^{2}} \sum_{i=1}^{n}
      W_{ij}
      \Bigg(\frac{2A_{i}-1}{A_{i}g(W_{i}) + (1-A_{i})(1-g(W_{i}))} \\
      & \qquad\qquad\qquad\qquad\qquad\quad \int_{0}^{t}
        \frac{I(\tilde{T}_{i} \geq u)}
        {c(u_{-}|A_{i},W_{i}) S(u|A_{i},W_{i})}
        \left(I(T_i=u)-\lambda(u|A_{i},W_{i})\right) du\\
      & \qquad\qquad\qquad\qquad\qquad\qquad +
      \log \frac{S(t|1,W_{i})}{S(t|0,W_{i})} \Bigg) \; .
  \end{split}
\end{equation*}

{\em Targeted maximum likelihood estimator.} This estimator employs a
conditional hazard estimator tilting procedure similar to that of
$\Psi_{j}^{(\text{TML})}(\hat{P}_n;t)$. The definition of $H_{j}(t|A,W)$ is slightly
modified:
\begin{equation*}
  H_{j}(t|A,W) \equiv \frac{W_{j}(2A-1)}
    {\left(Ag(W) + (1-A)(1-g(W))\right)\mathbb{E}_{P}\left[W_{j}^{2}\right]}
    \int_{0}^{t}
    \frac{1}{c(u_{-}|A,W)S(u|A,W)}
    du \;.
\end{equation*}
Then, given the tilted empirical distribution $P_{n}^{\star}$,
$\Gamma_{j}^{(\text{TML})}(\hat{P}_n;t) \equiv \Gamma_{j}(P_{n}^{\star}; t)$.

\paragraph{Asymptotic Behavior}

\begin{proposition}\label{prop:dr-tte-rel-vip-est}

  If A\ref{ass:centered-confounders}, A\ref{ass:non-zero-variance}, and
  A\ref{ass:cond-surv-consistency} are satisfied,
  $\Gamma^{(\text{OS})}(\hat{P}_n; t) \overset{P}{\rightarrow} \Gamma(P_{0}; t)$.
  The estimating equation and TML estimators share this property, too.
\end{proposition}

\begin{assumption}\label{ass:consistency-survival}
  Convergence rate of the conditional survival estimator:
  $\lVert S_{n}(t|A,W)-S_{0}(t|A,W)\rVert_{2} = o_{P}(n^{-1/4})$.
\end{assumption}

\begin{theorem}\label{thm:asymp-tte-rel-risks-ests}
  Assuming that A\ref{ass:centered-confounders}, A\ref{ass:non-zero-variance},
  A\ref{ass:donsker}, A\ref{ass:bounded-confounders},
  A\ref{ass:consistency-treatment-survival}, A\ref{ass:consistency-censoring},
  and A\ref{ass:consistency-survival}, are met,
  $\sqrt{n}(\Gamma_{j}^{(\text{OS})}(\hat{P}_n; t) - \Gamma_{j}(P_{0}; t)) = (1/\sqrt{n}) \sum_i D_j(O_{i}, P_{0}) + o_{P}(1)$.
  It follows that
  $\sqrt{n}(\Gamma_{j}^{(\text{OS})}(\hat{P}_n; t) - \Gamma_{j}(P_{0}; t)) \overset{D}{\rightarrow} N(0, \mathbb{E}_{P_{0}}[D_{j}(O, P_{0}; t)^{2}])$.
  This result applies to the estimating equation and TML estimators as well.
\end{theorem}

As for the relative TEM-VIP introduced in Equation~\eqref{eq:cont-rel-vip}, the
nonparametric estimators of the estimand in Equation~\eqref{eq:tte-rel-vip} are
not double-robust. Further, consistent estimation of all nuisance parameters at
the typical nonparametric rate is required to ensure the asymptotic linearity of
the estimators. For example, in an RCT where censoring is assumed to be
completely at random, consistent estimation of the survival function is
necessary to produce consistent estimates of $\Gamma(P_{0};t)$. If the
conditions of Theorem~\ref{thm:asymp-tte-rel-risks-ests} are satisfied, however,
then asymptotically valid hypothesis testing about the parameter is possible
using the Gaussian null distribution.

\section{Deriving New Treatment Effect Modification Variable Importance Parameters}\label{sec:extending-framework}

Readers might find the previous sections repetitive. This is purposeful. Their
contents provide a blueprint for defining pathwise differentiable TEM-VIPs based
on causal parameters of treatment effects, deriving estimators of these
TEM-VIPs, and establishing conditions under which these estimators are regular
and asymptotically linear and efficient. We formalize this framework in the
following workflow.

\begin{enumerate}
  \item Select a full-data, pathwise differentiable parameter $\Phi^{F}(P_{X})$
        of some treatment effect that is relevant to the problem at hand. For
        example, we consider the average treatment effect
        $\mathbb{E}_{P_{X}}[Y^{(1)}-Y^{(0)}]$ in
        Section~\ref{subsec:cont-abs-vip} for continuous and binary
        outcomes, and the difference in RMSTs
        $\mathbb{E}_{P_{X}}[\min\{T^{(1)}, t\} - \min\{T^{(0)}, t\}]$ in
        Section~\ref{subsec:tte-abs-vip} for right-censored time-to-event
        outcomes.
  \item Define $f(W)$ such that $\mathbb{E}_{P_{X}}[f(W)] = \Phi^{F}(P_{X})$.
        Under A\ref{ass:centered-confounders} and A\ref{ass:non-zero-variance},
        the TEM-VIP of covariate $j$ is given by
        $\Theta^{F}_{j}(P_{X}) = \mathbb{E}_{P_{X}}[f(W)W_{j}]/\mathbb{E}_{P_{X}}[W_{j}^{2}]$.
        In Section~\ref{subsec:cont-abs-vip},
        $f(W) = \bar{Q}_{P_{X}}(1, W) - \bar{Q}_{P_{X}}(0,W)$, the CATE, and in
        Section~\ref{subsec:tte-abs-vip},
        $f(W) = \int_{0}^{t} S_{P_{X}}(u|1, W) - S_{P_{X}}(u|0, W) \; du$, the
        conditional RMST. Note that $f(W)$ and $\Phi^{F}(P_{X})$ have a
          many-to-one relationship; the resulting $\Phi^{F}(P_{X})$ is not
          unique to a given $f(W)$.
  \item Establish the identifiability of the TEM-VIP in the observed-data model.
       Denoting the observed-data counterparts of
        $\Theta^{F}_{j}$ and $\Phi^{F}$ as $\Theta_{j}$ and $\Phi$,
        respectively, the conditions establishing that
        $\Theta^{F}_{j}(P_{X}) = \Theta_{j}(P)$ are virtually identical to the
        conditions needed for the equality of $\Phi^{F}(P_{X})$ and $\Phi(P)$.
        The only additional assumption required is that $W_{j}$ have bounded
        variance. See Theorems~\ref{thm:id-cond-abs-risk} and
        \ref{thm:id-tte-abs-vip} for examples.
  \item Derive the efficient influence function of the TEM-VIP. This derivation
        is straightforward, relying on the chain rule and the definition of the
        efficient influence function for $\Phi(P)$. If the uncentered efficient
        influence function of $\Phi(P)$ is given by $d(O, P)$ for $O\sim P$,
        then the efficient influence function of the TEM-VIP, $\Theta_{j}(P)$,
        based on $\Phi(P)$ is
        $W_{j}/\mathbb{E}_{P}[W_{j}^{2}](d(O, P) - W_{j}\Theta_{j}(P))$.
        Consider the average treatment effect, whose uncentered efficient
        influence function is
        $d(O,P) = (2A-1)/(Ag(A)+(1-A)(1-g(A)))(Y-\bar{Q}(A,W)) + \bar{Q}(1,W) - \bar{Q}(0,W)$.
        Using the previous formula, the efficient influence function of the
        absolute TEM-VIP with a continuous or binary outcome is that of
        Proposition~\ref{prop:eif-cont-abs-vip}, where the generic $\Theta_{j}$
        is replaced by $\Psi_{j}$. Similarly, the uncentered efficient influence
        function of the log-ratio of expected conditional potential outcomes is
        given by
        $d(O,P) = (2A-1)/(Ag(A)+(1-A)(1-g(A)))(Y-\bar{Q}(A,W))/\bar{Q}(A,W) + \log(\bar{Q}(1,W) / \bar{Q}(0,W))$.
        The efficient influence function of the relative TEM-VIP for continuous
        and binary outcome settings is given in
        Proposition~\ref{prop:eif-cont-rel-vip}, where
        $\Theta_{j} = \Gamma_{j}$.
  \item The one-step, estimating equation, and TML estimators can then be
        derived from the TEM-VIPs efficient influence function. Examples are
        found in Sections~\ref{sec:cont-outcome} and \ref{sec:tte-outcome}.
  \item These estimators' asymptotic properties are identical to those of the
        nonparametric efficient estimators of $\Phi$, like
        double-robustness, assuming that the potential treatment effect
        modifiers are bounded. Again, examples are provided in
        Sections~\ref{sec:cont-outcome} and \ref{sec:tte-outcome}.
\end{enumerate}

\section{Simulation Studies}\label{sec:simulations}

Next, we investigate the finite-sample performance of the proposed one-step and
TML estimators for a subset of the previously introduced estimands. Recall that
the one-step estimator is obtained by subtracting the empirical EIF from the
plug-in estimator --- and is equal, in the settings considered here, to the
estimating equation estimator --- and that the TML estimator is derived by first
tilting the nuisance parameter estimators to ensure that the mean of the
empirical EIF is negligible, and then using these updated estimators in the
plug-in estimator. The one-step and TML estimators are implemented in the
\texttt{unihtee} \texttt{R} software package, available
at
\href{https://github.com/insightsengineering/unihtee}{github.com/insightsengineering/unihtee}
and to be submitted to the Comprehensive R Archive Network (CRAN)
\citep{rstats}. These estimators' empirical absolute bias, variance, and Type I
error rates are evaluated in two observational study scenarios --- one with a
continuous outcome and another with a binary outcome --- and one RCT setting
with a time-to-event outcome. These simulation experiments rely on the
\texttt{simChef} \texttt{R} package's simulation study framework
\citep{simChef}. Code for reproducing these simulations is made available
at
\href{https://github.com/PhilBoileau/pub\_temvip-framework}{github.com/PhilBoileau/pub\_temvip-framework}.

The nonparametric estimators' capacity to recover treatment effect modifiers is
compared to that of \citet{tian2014} and \citet{chen2017}'s (augmented) modified
covariates methods. These methods are among the few that enable treatment effect
modification discovery in high-dimensional data under a variety of DGPs ---
albeit requiring stringent assumptions like sparsity and negligible correlation
structure among pre-treatment covariates. We stress, however, that their primary
goal is not the recovery of these treatment effect modifiers, but CATE
estimation. The modified covariates approach estimates the CATE by cleverly
transforming the outcome such that only the treatment-covariate interactions in
a GLM need be modeled. The augmented modified covariates procedure models this
transformed outcome as a function of all covariates to improve efficiency. Both
procedures can incorporate propensity score weights to improve estimation in
observational study scenarios. TEM discovery is possible when employing modeling
strategies with built-in feature selection capabilities; we model the
transformed outcome-covariates relationships with a linear model and fit this
model with the LASSO \citep{tibshirani1996}. Variables are classified as TEMs
when their estimated treatment-covariate interaction coefficients are non-zero.
These methods are implemented in the \texttt{personalized} R package
\citep{huling2021}.

\subsection{Continuous Outcome, Observational Study}

The first DGP we consider has a continuous outcome $Y$, high-dimensional
covariates $W$, and mimics an observational study, in that treatment status $A$
is an unknown function of $W$:
\begin{equation*}
  \begin{split}
    W & \sim N(0, I_{500 \times 500}) \\
    A|W & \sim \text{Bernoulli}\left(\text{logit}^{-1}
          \left(\frac{1}{4}\left( W_{1} - W_{2} + W_{3}\right)\right)\right) \\
    Y|A,W & \sim 1 + 2\left\lvert \sum_{j = 1}^{5} W_{j} \right\rvert +
            \left(5A - 2\right) \sum_{j=1}^{5} W_{j} + \epsilon \; ,
  \end{split}
\end{equation*}
where $\epsilon \sim N(0, 1/2)$. Note that the treatment assignment mechanisms
used here and in the following subsections were chosen to respect Assumption
A\ref{ass:positivity}. Indeed, the estimators --- particularly the TML
estimators --- presented in Sections~\ref{sec:cont-outcome} and
\ref{sec:tte-outcome} exhibit extreme variability in the presence of
\textit{random positivity violations}~\citep[][Fine Point
12.2]{hernan2023causal}. Random positivity violations materialize in finite
samples when the estimated probability of receiving treatment is negligible,
and can occur even when the positivity assumption of A\ref{ass:positivity} is
satisfied.

We take as target of inference the absolute TEM-VIPs of
Equation~\eqref{eq:cont-abs-vip}. We consider five sample sizes:
$n=125, 250, 500, 1,\!000,$ and $2,\!000$. Two hundred replicates are simulated
at each sample size.

We consider the one-step and TML estimators of this parameter where
$\bar{Q}_{0}$ and $g_{0}$ are estimated using the Super Learner algorithm of
\citet{vdl2007} implemented in the \texttt{sl3} R package \citep{coyle2021sl3}.
This algorithm computes the convex combination of nuisance parameter estimators,
referred to as base learners, that optimizes the cross-validated risk for the
squared error and negative log-likelihood loss function for the conditional
outcome and propensity score, respectively. For $\bar{Q}_{0}$, the base learners
are comprised of the LASSO \citep{tibshirani1996}, ridge regression
\citep{hoerl1970}, elastic net \citep{zou2005}, and multivariate adaptive
regression splines (MARS) \citep{friedman1991} estimators with main and
treatment-covariate interaction terms, as well as Random Forests
\citep{breiman2001}. For $g_{0}$, we consider the LASSO, ridge regression,
elastic net, MARS, and Random Forests. The modified covariates method and its
augmented counterpart estimate $g_{0}$ using LASSO, and employ the identity link
function to estimate the association of the covariates and treatment on the
outcome.

\begin{figure}
  \centering
  \includegraphics[width=1\textwidth]{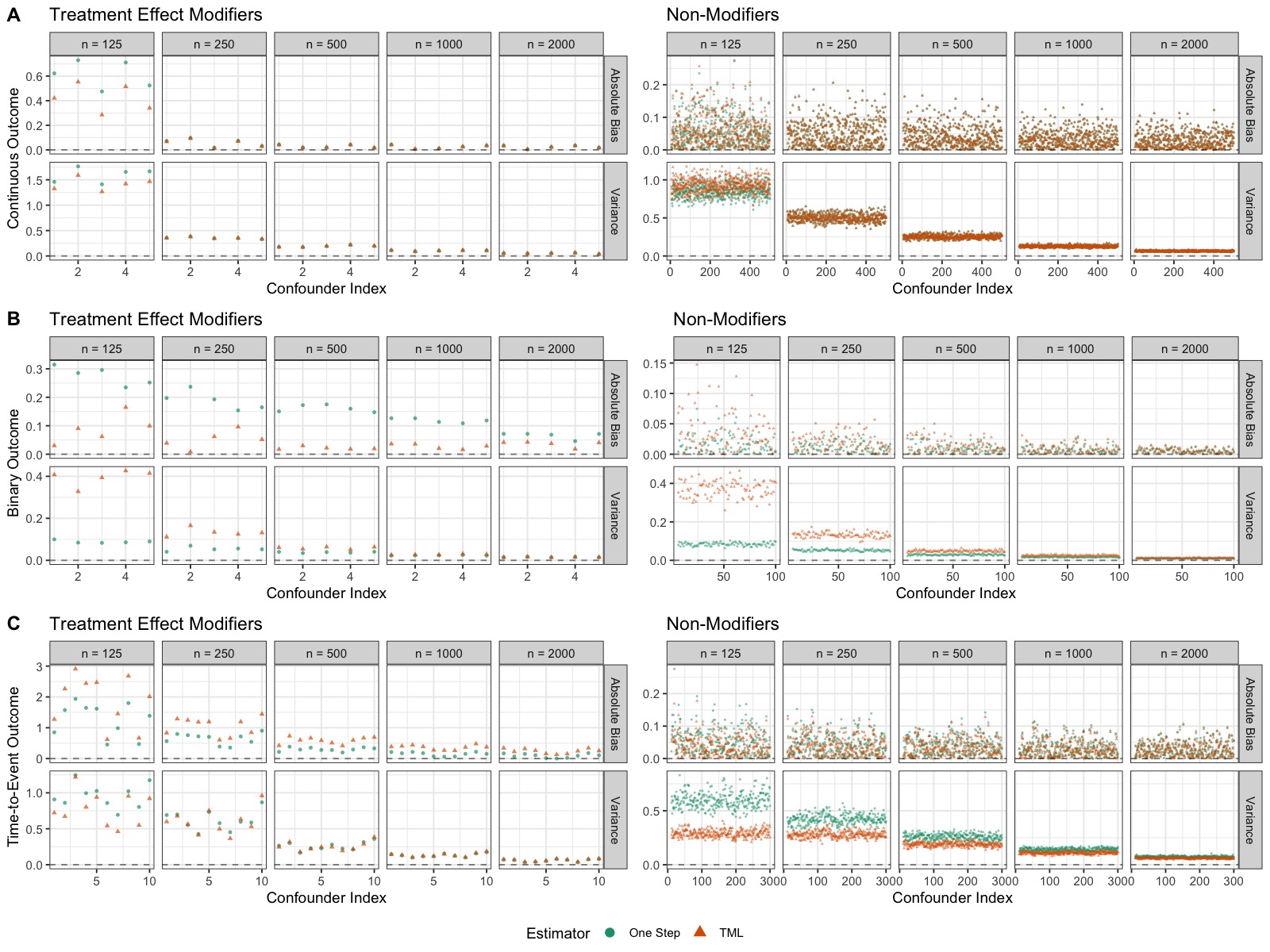}
  \caption{{\em Empirical bias and variance of one-step and TML estimators.} The
    empirical bias and variance of the one-step and TML estimators are
    stratified by DGP, treatment modifier status, and sample size (note the
    difference in y-axis scales between modifiers and non-modifiers). Two hundred
    replicates were simulated to compute the values in each scenario.}
  \label{fig:emp-bias-var}
\end{figure}

Figure~\ref{fig:emp-bias-var}A presents the empirical absolute bias and variance
of the one-step and TML estimators. Both exhibit a small empirical bias for the
TEMs for $n=125$, but are otherwise approximately unbiased at all other sample
sizes. These estimators' variances are virtually identical at all sample sizes,
and rapidly decrease as sample size increases. The bias and variance for
non-TEMs (covariates indices 6 to 500) are similarly negligible for both
estimators in all sample sizes.

\begin{figure}
  \centering
  \includegraphics[width=1\textwidth]{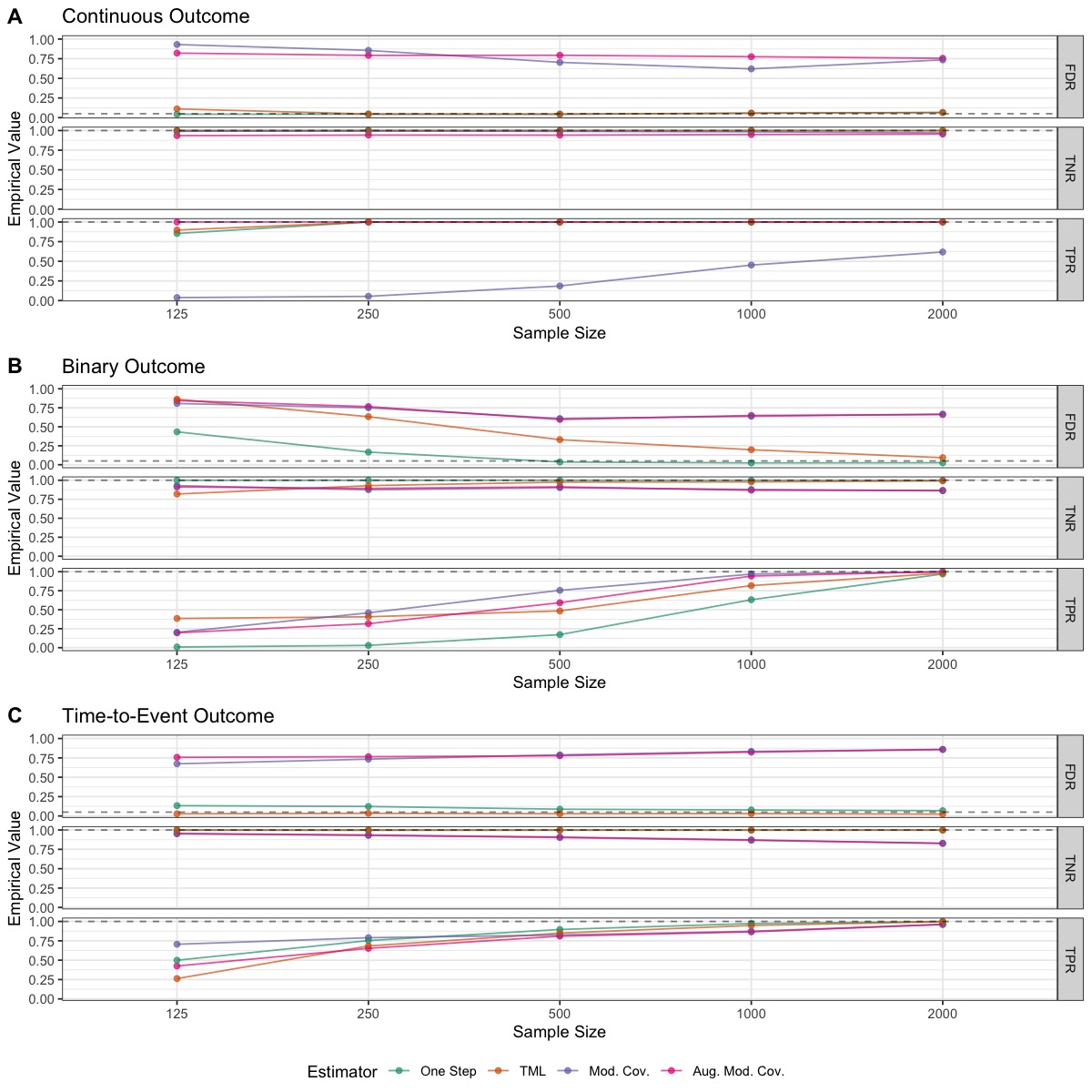}
  \caption{{\em TEM classification results.} The one-step, TML, modified
    covariates, and augmented modified covariates estimators' capacities to
    correctly identify TEMs from the set of covariates are measured in terms
    of the FDR, TPR, and TNR. These metrics are stratified by DGP and sample
    size. Two hundred replicates were simulated to compute the values in each
    scenario.}
  \label{fig:classification}
\end{figure}

We next evaluate these estimators' ability to distinguish covariates that modify
the effect of treatment from those that do not. The empirical false discovery
rate (FDR), true negative rate (TNR), and true positive rate (TPR) are computed
at each sample size. The FDR reports the proportion of incorrectly classified
covariates among the set of predicted TEMs. The TNR and TPR measure the
proportion of correctly classified non-TEMs and TEMs, respectively. Using
nominal 5\%-level, two-sided Wald-type hypothesis tests and accounting for
multiple testing using the FDR-controlling approach of \citet{benjamini1995}, we
expect the one-step and TML estimators to achieve a 5\% FDR in the largest
sample sizes. The one-step and TML estimators' classification are compared to
those of the modified covariates and augmented modified covariates methods.
Again, variables with non-zero estimated treatment-covariate interaction
coefficients are labeled as TEMs.

Of the four methods considered, only the one-step and TML estimators
approximately control the FDR at the nominal level in all sample sizes
(Figure~\ref{fig:classification}A). The (augmented) modified covariates methods,
on the other hand, maintain an FDR near 75\%. Their performance does not improve
as a function of $n$. Trends in the methods' FDRs are elucidated by their TNRs
and TPRs. The one-step and TML estimators produce a near-perfect TNR while
maintaining a competitive TPR. The augmented modified covariates procedure has a
TPR near 100\% in all sample sizes, yet has TNRs marginally lower than the
one-step and TML estimators. The modified covariates method produces similar
TNRs to its augmented counterpart, but has poorer TPRs. The parametric methods'
inability to reliably classify TEMs might be due to the non-linearity of the
expected conditional outcome or the number of features relative to the sample
size.

\subsection{Binary Outcome, Observational Study}

We consider another observational DGP, this time with a binary outcome and a
moderate number of correlated covariates:
\begin{equation*}
  \begin{split}
    W & \sim N(0, \Sigma_{100 \times 100}),
    \: \Sigma_{ij} = \begin{cases}
        1, & i = j \\
        0.1 \lvert i - j \rvert^{-1.8}, & \text{otherwise}
      \end{cases}\\
    A|W & \sim \text{Bernoulli}\left(\text{logit}^{-1}
          \left(\frac{1}{4} \left(W_{1} + W_{2} + W_{3}\right)\right)\right) \\
    Y|A,W & \sim \text{Bernoulli}\left(
            \text{logit}^{-1}\left(
            1 - 2A + \sum_{j=1}^{5} W_{j} + \left(A - \frac{1}{2}\right)
            \sum_{j=1}^{5} W_{j} \right)\right) \; .
  \end{split}
\end{equation*}
Here, $\Sigma$ is a $100 \times 100$ Toeplitz matrix, so that the pre-treatment
covariates' correlation structure imitates that of spatial or temporal data.
Again, care is taken to avoid random positivity violation issues.

We benchmark the estimation of the relative TEM-VIP presented in
Equation~\eqref{eq:cont-rel-vip}. The true parameter values are approximated
using Monte Carlo methods. Again, 200 replicates were simulated for each of
$n=125, 250, 500, 1,\!000,$ and $2,\!000$. The corresponding one-step and TML
estimators are compared and their nuisance parameters are estimated using Super
Learners, with the same base learners for $\bar{Q}_{0}$ and $g_{0}$ as in the
continuous-outcome example. The (augmented) modified covariate methods again
rely on the LASSO for propensity score estimation, and use the logistic link
function to model the outcome conditional on the potential TEMs and treatment.

The empirical bias and variance of the one-step and TML estimators are provided
in Figure~\ref{fig:emp-bias-var}B. Among the TEMs, the one-step estimator
exhibits more finite-sample bias than the TML estimator, though this bias
decreases as sample size increases. The TML estimator, however, has noticeably
greater finite-sample variance than the one-step for $n=125, 250$. Among the
non-TEMs (pre-treatment covariates indexed 6 through 100), these estimators have
similar bias. Again, however, the TML estimator has greater variance in the
smaller sample sizes.

The empirical FDR, TNR, and TPR of the one-step and TML estimators, as well as
those of the (augmented) modified covariates methods are presented in
Figure~\ref{fig:classification}B. Only the one-step estimator reliably controls
the FDR at the 5\% level at sample sizes of $500$ and above. This is seemingly
due to the estimator's conservative behavior: It achieves a near-perfect TNR at
all sample sizes, but has the lowest TPR of all estimators regardless of sample
size. The TML estimator fails to control the FDR at the desired levels in all
sample sizes, though the FDR decreases with sample size and is nearly controlled
at $n=2\!,000$. The poor FDR of the TML estimator relative to the one-step
estimator may be due to the latter's increased variability, exhibited in
Figure~\ref{fig:emp-bias-var}B. The (augmented) modified covariates methods tend
to perform similarly: their FDR hovers around 75\% at all sample sizes, their
TNR decreases marginally as $n$ increases, and their TPRs are generally higher
than those of the nonparametric estimators. Given that sparsity and linearity
assumptions are satisfied, the lackluster FDR control of the (augmented)
modified covariates procedures might be attributed to violations of the
Irrepresentable Condition \citep{zhao2006} --- the covariates' correlation
structure is too complex.

\subsection{Right-Censored Time-to-Event Outcome, Randomized Control Trial}

Next, we simulate RCT data with known treatment assignment mechanism, a discrete
right-censored time-to-event outcome, and a duration of 10 time units. Recall
that $O = (W, A, \tilde{T}, \Delta)$, where $W$ and $A$ are defined as before,
$\tilde{T}$ is the right-censored time-to-event, and $\Delta$ is the censoring
indicator. The simulation generative model is given by
\begin{equation*}
  \begin{split}
    W & \sim N(0, \Sigma_{300 \times 300}) \\
    A & \sim \text{Bernoulli}(1/2) \\
    C|A,W & \sim \min\left\{\text{Negative Binomial}\left(1,
                  \text{logit}^{-1}\left(5 + A + W_{1}\right)\right),
                10 \right\} \\
    T|A,W,C & \sim \text{Negative Binomial}\left(1,
            \text{logit}^{-1}\left(
              -2 - A + (10A - 5) \sum_{j=1}^{10} W_{j}
            \right)\right) \\
    \tilde{T} & = \min\left\{T, C \right\}\\
    \Delta & = I(T > c),
  \end{split}
\end{equation*}
where the covariates' covariance matrix $\Sigma$ is block-diagonal, with each
block corresponding to ten moderately correlated features. This correlation
structure loosely mimics the expression levels of a collection of
genes.

The estimand is defined as the absolute TEM-VIP of
Equation~\eqref{eq:tte-abs-vip} at time $t=9$. Again, the true parameter values
are approximated through Monte Carlo methods. The one-step and TML estimators'
conditional censoring hazard function is estimated by the LASSO and their
conditional survival hazard function is estimated by the LASSO augmented with
treatment-covariate interaction terms. The propensity scores of these
nonparametric estimators and the (augmented) modified covariates methods are
fixed at $1/2$, as in a 1:1 RCT. Penalized Cox proportional hazards models are
used by the parametric methods to model the conditional survival hazard. We
highlight that our simulation DGP satisfies the proportional hazards and
non-informative censoring assumptions, but that its covariates possess a complex
correlation structure. This might worsen the (augmented) modified covariate
methods' treatment effect modifier classification performance.

Figure~\ref{fig:emp-bias-var}C presents the one-step and TML estimators'
empirical biases and variances. As for the binary DGP, both estimators are
biased for the TEMs (indices 1--10) at all sample sizes, but approximately
unbiased for all non-TEMs. As expected, however, the empirical bias associated
with the TEMs decreases with sample size, and is negligible when $n=2,\!000$.
The empirical variances of these estimators behave as expected, too: they
decrease with increasing sample size. The TML estimator's empirical variances
are generally smaller than those of the one-step estimator.

The FDR, TNR, and TPR of all methods considered are reported in
Figure~\ref{fig:classification}C. The TML estimator is the only procedure to
control the FDR at the nominal 5\% level, while the one-step estimator possesses
an FDR of approximately 10\% for $n = 125, 250$, and which slowly decreases to
the nominal rate by $n=2,\!000$. The (augmented) modified covariates approaches
result in empirical FDRs that grow with sample size, from approximately 70\% for
$n=125$ to 90\% for $n=2,\!000$. The parametric methods' behavior with respect
to the FDR might be explained by the relationship between their TNR and sample
size: as sample size increases, they produce a greater amount of false
positives. The nonparametric estimators, however, maintain a near-perfect TNR at
all sample sizes. All procedures perform similarly with respect to the TPRs in
all but the smallest sample size.

\section{Application}\label{sec:rct-data-application}

We apply our framework to a clinical trial dataset with a right-censored
time-to-event outcome. This analysis, as well as the results of the simulation
studies, can be reproduced with the code found in this public repository:
\href{https://github.com/PhilBoileau/pub\_temvip-framework}{github.com/PhilBoileau/pub\_temvip-framewor}.

Trastuzumab is a monoclonal antibody targeting the \textit{HER2} oncogene that
demonstrably improves the clinical outcomes of breast cancer patients whose
tumors over-express this gene. Improvement is not uniform, however: some
patients are resistant to this therapy. Identifying biomarkers that predict
response to trastuzumab is therefore of great interest \citep{loi2014}.

\citet{loi2014} make available a subset of patients enrolled in the FinHER
clinical trial (GSE47994), a study comparing docetaxel and vinorelbine ---
chemotherapies --- as adjuvant treatment for early-stage breast cancer
\citep{joensuu2006}. Patients with over-expressed \textit{HER2} disease were
additionally randomized to receive either nine weekly trastuzumab infusions or
no trastuzumab . \citet{loi2014} provide the quality controlled, normalized gene
expression data and relevant clinical information for 201 of these patients.
Taking as outcome distant disease-free survival, defined as the time interval
between the date of randomization and the date of first cancer recurrence or
death, if prior to recurrence, we consider the 500 most variable genes for the
purpose of TEM discovery.

Traditional approaches to this task rely on Cox proportional hazards models. For
example, a penalized regression of the outcome on the treatment, genes,
treatment-gene interactions, and pre-treatment covariates like age and
chemotherapy could be fit, and the genes with non-zero estimated interaction
coefficients would be classified as TEMs. This is similar to the augmented
modified covariates approach of \citet{tian2014}. Alternatively, individual
regressions for each gene of the outcome conditioning on treatment, gene,
pre-treatment covariates, and the treatment-gene interaction could be fitted.
Genes with significant treatment-gene interactions would be reported as TEMs.
However, both approaches perform inference about conditional parameters, the
hazards ratio, while we aim to learn about parameters that reflect
population-level information about treatment effect heterogeneity. Verifying the
proportional hazards assumption is also impractical given the number of
potential TEMs considered.

We instead use our framework, taking as estimand the RMST-based TEM-VIP of
Equation~\eqref{eq:tte-abs-vip}. Patients' distant disease-free survival times
are discretized into 6-month intervals for computational convenience. We use the
TML estimator since the previous simulation experiments suggests that it
controls the Type I error rate better than the one-step estimator at this sample
size. Its element-wise variance is also likely lower. Given that previous
evidence suggests possible higher-order interactions between patients'
chemotherapy regimen, trastuzumab, and biomarkers \citep{loi2014}, we estimate
the conditional failure and censoring hazards using a Super Learner made up of
the penalized generalized linear models using the logit link and possessing
terms for the treatment, genes, and treatment-gene interactions, Random Forests,
and XGBoost \citep{chen2016}. This procedures takes approximately 20 minutes to
run on a personal computer with a single core of an Apple M1 CPU.
Parallelization can reduce this runtime further. We note that similar results
are produced by directly estimating the nuisance parameters with Random Forests
or XGBoost, though at the expense of an objective choice of nuisance estimators
otherwise facilitated by the Super Learner estimator.

\begin{figure}
  \centering
  \includegraphics[width=1\textwidth]{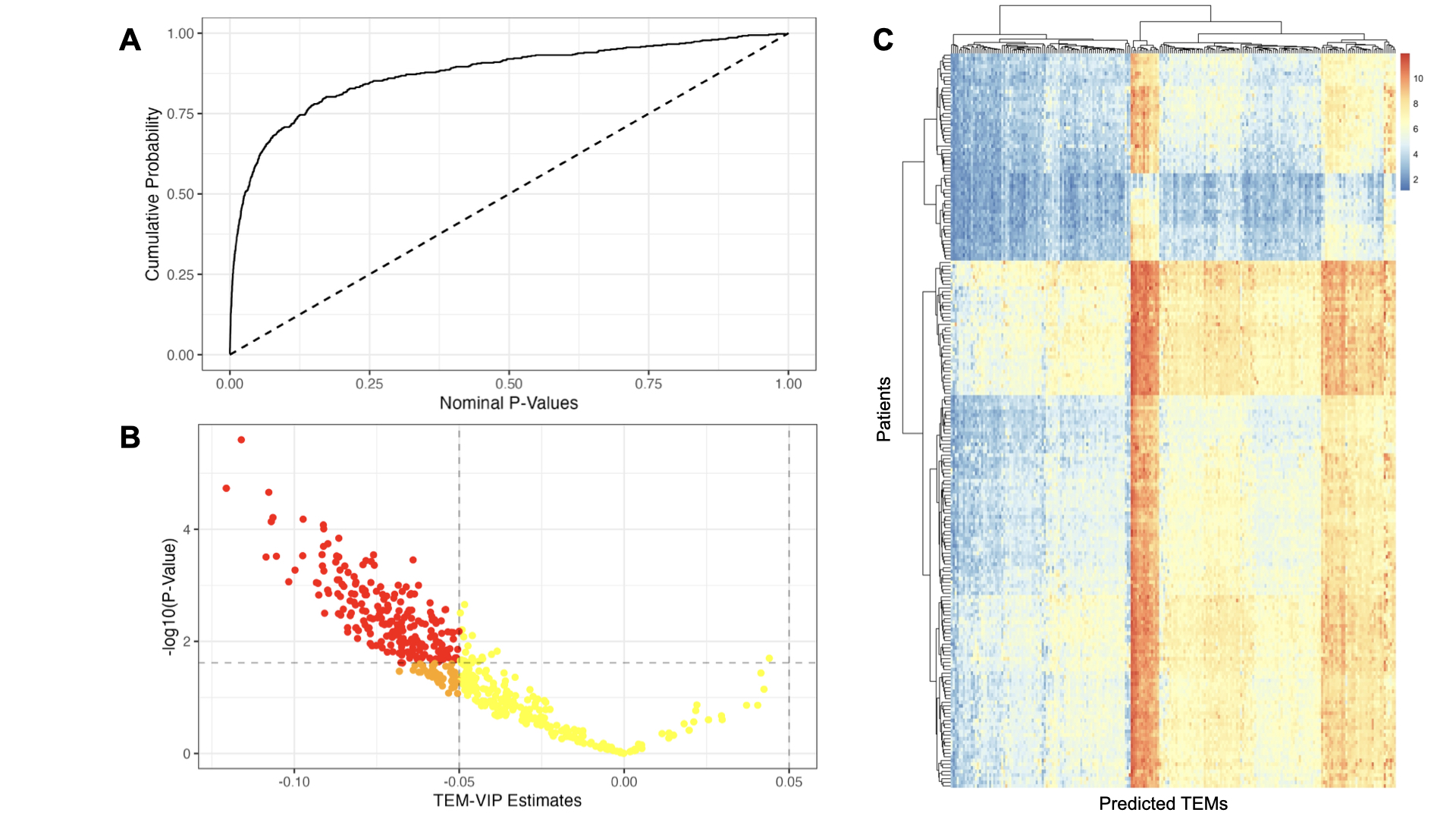}
  \caption{{\em FinHER clinical trial data analysis results.} \textbf{A}
    Empirical cumulative distribution function (eCDF) of nominal $p$-values. The
    dotted line corresponds to the eCDF under the null (a
    $\text{Uniform}([0,1])$ distribution). \textbf{B} Volcano plot of the 500
    most variable genes' TEM-VIP estimates and associated nominal $p$-values.
    Yellow genes are deemed unimportant due to their small estimated effect
    sizes and larger $p$-values; orange genes possess a meaningful estimated
    effect but fail to achieve the adjusted $p$-value cutoff; red genes are
    significant at the 5\% FDR level and have large estimated TEM-VIPs.
    \textbf{C} The log-transformed gene expression data of genes with meaningful
    effect estimates are used to cluster patients. Hierarchical clustering with
    complete linkage is used for patients and identified TEMs alike.}
  \label{fig:gene-expression}
\end{figure}

In this analysis, we have sought to dichotomize pre-treatment covariates into
TEMs and non-TEMs based on the value of their estimated TEM-VIP. However, as
expected, there seems to be a continuum in the biomarkers' capacity to influence
the effect of treatment, in terms of both statistical significance and
biological effect size. This can be seen in the empirical cumulative
distribution function (eCDF) of the nominal $p$-values
(Figure~\ref{fig:gene-expression}A) and in the volcano plot
(Figure~\ref{fig:gene-expression}B). Hypothesis testing alone, with a null of
$\Psi(P_{0}) = 0$, may therefore not be adequate. As in differential expression
studies in transcriptomics, one can instead leverage the volcano plot and deem a
biomarker of clinical interest if it is significant at the 5\% FDR level
\textit{and} if its absolute estimated TEM-VIP is larger than $0.05$ (for each
unit increase in log2 gene expression, a TEM-VIP equal to $0.05$ in this
analysis approximately corresponds to an expected difference in RMST of about
$18$ days). There are 220 such biomarkers for the FinHER clinical trial.
Alternatively, if one is interested only in modifications above a certain
magnitude $m$, one could define the null hypothesis for the $j^{\text{th}}$
biomarker as $\lvert \Psi_{j}(P_{0}) \rvert \leq m$. The (adjusted) $p$-values
obtained from these tests could then be used to produce a ranked list of
biomarkers for follow-up analyses. The above considerations highlight the
importance of thinking carefully and critically about how to translate the
biological question of interest into a statistical inference question, including
defining what constitutes a meaningful effect size.

\begin{table}
  \centering
  \begin{minipage}{5cm}
      \caption{Top five selected TEMs}\label{tab:top-five-tems}
      \begin{tabular}{rlrrr}
        \hline
        & Gene & Estimate & Std. Err. & Adj. $p$-Value \\
        \hline
        1 & EPPK1 & -0.116 & 0.025 & 0.001 \\
        2 & NDUFB3 & -0.121 & 0.028 & 0.004 \\
        3 & BNIP3L & -0.108 & 0.025 & 0.004 \\
        4 & PNKD & -0.106 & 0.027 & 0.006 \\
        5 & DUSP4 & -0.097 & 0.024 & 0.006 \\
        \hline
      \end{tabular}
  \end{minipage}
\end{table}

Now, the five genes with the smallest $p$-values from among the clinically
meaningful biomarkers are presented in Table~\ref{tab:top-five-tems}. All have
previously been linked to breast cancer, and their estimated effects are
generally in the direction expected by the literature. Increased \textit{EPPK1}
expression has been linked to estrogen-related receptor $\gamma$, which is
associated with breast cancer growth suppression \citep{ariazi2002,tiraby2011}.
A meta-analysis of 11 genome-wide association studies found that a single
nucleotide polymorphism in a \textit{NDUFB3} promoter was significantly
associated estrogen receptor negative breast cancer \citep{couch2016}.
\citet{moussay2011} found that \textit{BNIP3L} upregulation is associated with
TNF$\alpha$ stimulation, which is associated with trastuzumab resistance
\citep{mercogliano2017}. Evidence suggests that overexpression of
\textit{MR-1S}, an isomer of \textit{PNKD} associated with disordered cell
differentiation, malignant transformation initiation, and accelerated
metastasis, is therefore a potential therapeutic target of breast cancer
\citep{wang2018}. Finally, \citet{menyhart2017} found that increased expression
of \textit{DUSP4} correlates with increased resistance to trastuzumab.

We also present the log-transformed gene expression of the features with
clinically meaningful TEM-VIP estimates in Figure~\ref{fig:gene-expression}C. We
should expect them to define patient subgroups if these biomarkers truly modify
the effect of treatment. Indeed, these genes' expression data produce multiple
distinct patient clusters. We refrain from interpreting
Figure~\ref{fig:gene-expression}C any further, however, considering it solely a
diagnostic tool. Using patients' outcomes and biomarkers to compute TEM-VIP
estimates, then relying on these estimates to data-adaptively define subgroups
in the same data may cause overfitting. These results would ideally be validated
on an external dataset, though, as is often the case with openly-accessible
clinical trial data, none are available. This might motivate extensions to this
TEM discovery framework that support valid inference about both TEM-VIPs and
patient subgroups using the same data.

\section{Discussion}\label{sec:discussion}

We propose several causally interpretable TEM-VIPs in full-data models,
establish identifiability conditions to relate them to parameters of
observed-data distributions, derive accompanying nonparametric estimators, and
study these estimators' asymptotic behavior. Under non-stringent conditions on
the DGPs and nuisance parameter estimators, we find that these estimators are
consistent. Imposing a few additional assumptions results in efficient,
asymptotically linear estimators that permit straightforward hypothesis testing
about the corresponding TEM-VIPs. A general workflow for creating new TEM-VIPs
and deriving associated nonparametric estimators is provided.

Simulation experiments demonstrate that the estimators' behavior approximates
their established theoretical guarantees in realistic DGPs and for moderate
sample sizes. As an additional validation of our methodology, we attempted to
identify TEMs in a publicly available clinical trial dataset. These data were
originally collected to assess the effect of a monoclonal antibody therapy,
trastuzumab, on breast cancer patients. Many genes were classified as TEMs, and
a literature review of the top-ranked genes suggests that they are associated
with breast cancer. Indeed, a number of these TEMs are known biomarkers of
trastuzumab resistance. A diagnostic plot of the predicted TEMs' expression data
further suggests that they may be used to define patient subgroups, but this
must be validated with external data.

This work gives rise to several research directions. The framework outlined in
Section~\ref{sec:extending-framework} permits the derivation of bespoke pathwise
differentiable TEM-VIPs and accompanying nonparametric efficient estimators. In
particular, researchers working in the biotechnology and pharmaceutical
industries can perform inference about TEM-VIPs derived from estimands used in
clinical trials. Such heterogeneous treatment effect analyses would closely
track the statistical guidelines enforced by regulatory authorities, like those
of the International Council for Harmonization of Technical Requirements for
Pharmaceuticals for Human Use for clinical trials \citep{ich-estimand}. This
framework for TEM inference might also support statistically rigorous subgroup
discovery. TEMs identified using our methodology could be used to cluster
observations (i.e., patients), and, subsequently, treatment effects could be
estimated within these groups. Whether there exists a sound approach that
permits the application of this workflow to a single dataset, perhaps building
on recent advances in post-selection inference, should be investigated. Future
work might also determine whether these TEM-VIP estimators improve treatment
rule estimation procedures by acting as variable filters. That is, only
pre-treatment covariates with TEM-VIP estimates significantly different from
zero would be used, along with known confounders, to learn the treatment rule.
Doing so would increase the interpretability of the rule and might improve
estimation in high-dimensional regimes. Finally, as discussed in
Section~\ref{subsec:cont-abs-vip}, the TEM-VIPs considered here are a function
of their respective covariates' variances. Like the coefficients of a linear
model, these TEM-VIPs may not be comparable when the covariates' variances
differ. While covariates can be scaled using their sample standard deviations
prior to any analyses, future work might instead propose and study standardized
TEM-VIPs.

\subsubsection*{Acknowledgments}

P.B. is grateful for the support of the FRQNT (B2X) and NSERC (PGS-D).

\section*{Appendix}

\renewcommand*{\thesubsection}{S\arabic{subsection}}
\setcounter{table}{0}
\renewcommand{\thetable}{S\arabic{table}}
\setcounter{figure}{0}
\renewcommand{\thefigure}{S\arabic{figure}}
\setcounter{equation}{0}
\renewcommand{\theequation}{S\arabic{equation}}

\subsection*{Theorem~\ref{thm:id-cond-abs-risk}}
\begin{proof}
  It follows immediately from A\ref{ass:no-unmeasured-confounding} and
  A\ref{ass:positivity} that $\bar{Q}_{P_{X,0}}(A,W) = \bar{Q}_{0}(A,W)$.
  Then $\Psi^{F}(P_{x,0}) = \Psi(P_{0})$.
\end{proof}

\subsection*{Proposition~\ref{prop:eif-cont-abs-vip}}
\begin{proof}
  Using the generic definition provided in
  Equation~\eqref{eq:generic-eif}, the efficient influence function of
  $\Psi_{j}(O,P)$ is
  \begin{equation*}
    \begin{split}
      D_j(O,P)
      & = \frac{d}{d\epsilon} \Psi_{j}(P_{\epsilon}) \big|_{\epsilon = 0} \\
      & = \frac{W_j}{\mathbb{E}_P\left[W_j^2\right]}
        \left(\frac{I(A=1)}{g(W)}\left(Y - \bar{Q}(A,W)\right)
        + \bar{Q}(1, W)\right. \\
      & \qquad\qquad\qquad\qquad\qquad \left.
        - \frac{I(A=0)}{1-g(W)}\left(Y - \bar{Q}(A,W)\right)
        - \bar{Q}(0, W) - \Psi_j(P)W_j\right) \;.
    \end{split}
  \end{equation*}
\end{proof}

\subsubsection*{Proposition~\ref{prop:dr-cont-abs-risk}}
\begin{proof}
  From the definition of $D_j$ given by
  Equation~\eqref{eq:eif-cont-abs-risk}, we find that
  \begin{equation*}
    \begin{split}
      \mathbb{E}_{P_{0}}\left[D_j(O,P)\right]
      & \propto \mathbb{E}_{P_{0}}\left\{W_j\left(\left(\frac{g_0(W)}{g(W)}-1\right)
        \left(\bar{Q}_0(1,W)-\bar{Q}(1,W)\right) \right. \right.\\
      & \qquad\qquad\qquad\left.\left. - \left(\frac{1-g_0(W)}{1-g(W)}-1\right)
        \left(\bar{Q}_0(0,W)-\bar{Q}(0,W)\right)\right)\right\} \; .
    \end{split}
  \end{equation*}
  It follows immediately that
  $\mathbb{E}_{P_0}[D_j(O,P)] = 0$ when $g = g_0$ or
  $\bar{Q} = \bar{Q}_0$ for $j = 1, \ldots, p$.
\end{proof}

\subsubsection*{Theorem~\ref{thm:asymp-cont-abs-risk-ests}}
\begin{proof}

  Asymptotic linearity of $\Psi^{(\text{OS})}$ and $\Psi^{(\text{TML})}$ are
  achieved when the third and fourth terms in the von Mises expansion of
  Equation~\eqref{eq:von-mises} converge in probability to 0. Under
  A\ref{ass:donsker}, $D(O, \hat{P}_{n}) \in \mathcal{G}_0$ with probability tending to
  one which implies that
  $\mathbb{E}_{P_0}[(D_j(O,\hat{P}_{n})- D_j(O,P_0))^2] = o_P(1)$. It follows that the
  third term of the von Mises expansion is $o_P(1)$.

  What remains is to bound the error term. From \citet{boileau2022}, we find
  that
  \begin{equation*}
    \begin{split}
      -R(P_0, \hat{P}_{n})
      & = \frac{1}{\mathbb{E}_{P_0}\left[W_{j}^2\right]}
      \mathbb{E}_{P_0}\left[W_{j}\left(
          \frac{g_0(1, W)}{g_n(1, W)} - 1\right)\left(\bar{Q}_0(1,
        W)-\bar{Q}_n(1, W)\right) \right.\\
      & \qquad\qquad\qquad\qquad \left. - W_{j}\left(
          \frac{g_0(0, W)}{g_n(0, W)} - 1\right)\left(\bar{Q}_0(0,
        W)-\bar{Q}_n(0, W)\right)\right] \\
      & \leq \frac{1}{\mathbb{E}_{P_0}\left[W_{j}^2\right]}
      \left(\left\lvert
      \mathbb{E}_{P_0}\left[W_{j}\left(
          \frac{g_0(1, W)}{g_n(1, W)} - 1\right)\left(\bar{Q}_0(1,
        W)-\bar{Q}_n(1, W)\right) \right]\right\rvert\right.\\
      & \qquad\qquad\qquad\qquad + \left.\left\lvert\mathbb{E}_{P_0}
        \left[W_{j}\left(
          \frac{g_0(0, W)}{g_n(0, W)} - 1\right)\left(\bar{Q}_0(0,
          W)-\bar{Q}_n(0, W)\right)\right]\right\rvert\right) \\
      & \leq \frac{1}{\mathbb{E}_{P_0}\left[W_{j}^2\right]}
        \left(
        \mathbb{E}_{P_0}\left[W_{j}^2\left(
        \frac{g_0(1, W) - g_n(1, W)}{g_n(1, W)}\right)^2 \right]^{1/2}
          \mathbb{E}_{P_0}\left[\left(\bar{Q}_0(1, W)-\bar{Q}_n(1, W)
          \right)^2\right]^{1/2}\right. \\
      & \qquad\qquad\qquad\qquad + \left.
        \mathbb{E}_{P_0}\left[W_{j}^2\left(
          \frac{g_0(1, W) - g_n(1, W)}{g_n(0, W)}\right)^2\right]^{1/2}
          \mathbb{E}_{P_0}\left[\left(\bar{Q}_0(0, W)-\bar{Q}_n(0,
        W)\right)^2\right]^{1/2}\right) \\
      & \overset{\text{a.s.}}{\leq} \frac{C^{2}}
        {\mathbb{E}_{P_{0}}\left[W_{j^{2}}\right]}\left(
        \mathbb{E}_{P_0}\left[
        \left(\frac{g_0(W)}{g_{n}(W)} - 1\right)^2\right]^{1/2}
        \mathbb{E}_{P_0}\left[\left(\bar{Q}_0(1,W)
        - \bar{Q}_{n}(1,W)\right)^{2}\right]^{1/2} \right.\\
      & \qquad\qquad\qquad\qquad + \left.\mathbb{E}_{P_0}\left[
        \left(\frac{1-g_0(W)}{1-g_{n}(W)} - 1\right)^2\right]^{1/2}
        \mathbb{E}_{P_0}\left[\left(\bar{Q}_0(0,W)
        - \bar{Q}_{n}(0,W)\right)^{2}\right]^{1/2}\right) \; .\\
    \end{split}
  \end{equation*}
  The last inequality follows from A\ref{ass:bounded-confounders}. A
  similar bound applies to $R(P_0, \hat{P}_{n})$. The remainder term of
  Equation~\eqref{eq:von-mises} is therefore $o_P(1)$ under the conditions
  of A\ref{ass:double-rate-robustness}.

  It follows, applying the central limit theorem to the first term of the von
  Mises expansion, that $\sqrt{n}(\Psi_j^{(\text{OS})}(\hat{P}_{n})-\Psi_j(P_0)
  \overset{D}{\rightarrow} N(0, P_0D_j(O,P_0))$. The same is true for
  $\Psi_j^{(\text{TMLE})}(\hat{P}_{n})$.

\end{proof}

\subsection*{Corollary~\ref{cor:id-cond-rel-risk}}
\begin{proof}
  The conditions outlined in Theorem~\ref{thm:id-cond-abs-risk} imply that
  $\bar{Q}_{P_{X,0}}(A,W) = \bar{Q}_{0}(A,W)$. It follows immediately that
  $\Gamma^{F}(P_{X,0})$ is equal to $\Gamma(P_{0})$.
\end{proof}

\subsection*{Proposition~\ref{prop:eif-cont-rel-vip}}
\begin{proof}
  Using the same point mass contamination approach, we obtain the following efficient
  influence function for $\Gamma_{j}(O,P)$:
  \begin{equation*}
    \begin{split}
      D_j(O,P)
      & = \frac{d}{d\epsilon} \Gamma_{j}(P_{\epsilon}) \big|_{\epsilon = 0} \\
      & = \frac{W_j}{\mathbb{E}_{P}\left[W_j^2\right]}
        \left(\frac{I(A=1)}{g(W)\bar{Q}(1,W)}\left(Y - \bar{Q}(A,W)\right)
        + \bar{Q}(1, W)\right. \\
      & \qquad\qquad\qquad\qquad\qquad \left.
        - \frac{I(A=0)}{(1-g(W))\bar{Q}(0,W)}\left(Y - \bar{Q}(A,W)\right)
        + \bar{Q}(0, W) - \Psi_j(P)W_j\right) \;.
    \end{split}
  \end{equation*}
\end{proof}

\subsubsection*{Proposition~\ref{prop:cons-cont-rel-risk}}
\begin{proof}
  From the definition of $D_j$ given by
  Equation~\eqref{eq:eif-cont-rel-vip}, we find that
  \begin{equation*}
    \begin{split}
      \mathbb{E}_{P_{0}}\left[D_j(O,P)\right]
      & \propto \mathbb{E}_{P_0}\left\{W_j\left(\frac{g_0(W)}{g(W)\bar{Q}(A,W)}
        \left(\bar{Q}_0(1,W)-\bar{Q}(1,W)\right) +
      \log\bar{Q}_0(1,W)-\log\bar{Q}(1,W) \right. \right.\\
      & \qquad\qquad\qquad\left. \left. -
        \frac{1-g_0(W)}{\left(1-g(W)\right)\bar{Q}(A,W)}
        \left(\bar{Q}_0(0,W)-\bar{Q}(0,W)\right) +
        \log\bar{Q}_0(0,W)-\log\bar{Q}(0,W)\right)\right\} \; .
    \end{split}
  \end{equation*}
  It follows immediately that $\mathbb{E}_{P_0}[D_j(O,P)] = 0$ when
  $\bar{Q} = \bar{Q}_0$.
\end{proof}

\subsubsection*{Theorem~\ref{thm:asymp-cont-rel-risk-ests}}
\begin{proof}
  The proof is analogous to Theorem~\ref{thm:asymp-cont-abs-risk-ests}.
  Again, the entropy constraint of A\ref{ass:donsker} ensures that the
  third term of the von Mises expansion converges to zero in probability to 1.
  The remainder term in the same von Mises expansion is shown to be
  $o_P(n^{-1/2})$:
  \begin{equation*}
    \begin{split}
      -R(P_0,\hat{P}_{n})
      & = \frac{1}{\mathbb{E}_{P_0}\left[W_{j}^2\right]}
        \mathbb{E}_{P_0}\left[W_{j}\left(
        \frac{g_0(W)}{g_{n}(W)\bar{Q}_{n}(A,W)}
        \left(\bar{Q}_0(1,W)-\bar{Q}_{n}(1,W)\right)
        + \log\bar{Q}_0(1, W) - \log\bar{Q}_{n}(1, W) \right.\right.\\
      & \qquad\qquad\qquad\qquad \left.\left.
          - \frac{1-g_0(W)}{\left(1-g_{n}(W)\right)\bar{Q}_{n}(A,W)}
          \left(\bar{Q}_0(0,W)-\bar{Q}_{n}(0,W)\right)
          - \log\bar{Q}_0(0, W) + \log\bar{Q}_{n}(0, W)\right)\right] \\
      & \propto \mathbb{E}_{P_0}\left[W_{j}
        \frac{g_0(W)}{g_{n}(W)\bar{Q}_{n}(A,W)}
        \left(\bar{Q}_0(1,W)-\bar{Q}_{n}(1,W)\right)\right]
        + \mathbb{E}_{P_0}\left[W_j\left(\log\bar{Q}_0(1, W)-
        \log\bar{Q}_{n}(1, W)\right)\right] \\
      & \qquad\qquad
        - \mathbb{E}_{P_0}\left[W_j\frac{1-g_0(W)}{\left(1-g_{n}(W)\right)
        \bar{Q}_{n}(A,W)} \left(\bar{Q}_0(0,W)-\bar{Q}_{n}(0,W)\right)\right] \\
      & \qquad\qquad\qquad- \mathbb{E}_{P_0}\left[W_j\left(\log\bar{Q}_{n}(0, W) +
        \log\bar{Q}_0(0, W)\right)\right] \\
      & = \mathbb{E}_{P_0}\left[W_{j} \frac{g_0(W)}{g_{n}(W)}
        \left(\frac{\bar{Q}_0(1,W)}{\bar{Q}_{n}(1,W)}- 1 \right)\right]
        - \mathbb{E}_{P_0}\left[W_j\left(\frac{\bar{Q}_0(1, W)}{\bar{Q}_{n}(1, W)}-1
        \right)\right] \\
      & \qquad\qquad
        - \mathbb{E}_{P_0}\left[W_j\frac{1-g_0(W)}{\left(1-g_{n}(W)\right)}
        \left(\frac{\bar{Q}_0(0,W)}{\bar{Q}_{n}(0,W)} - 1\right)\right]
        + \mathbb{E}_{P_0}\left[W_j\left(\frac{\bar{Q}_0(0, W)}{\bar{Q}_{n}(0, W)}-1
        \right)\right] + o_P(n^{-1/2})\\
      & \leq \left\lvert \mathbb{E}_{P_0}\left[W_{j}
        \left(\frac{g_0(W)}{g_{n}(W)} - 1 \right)
        \left(\frac{\bar{Q}_0(1,W)}{\bar{Q}_{n}(1,W)} - 1\right)\right]\right\rvert\\
      & \qquad\qquad
        + \left\lvert\mathbb{E}_{P_0}\left[W_j
        \left(\frac{1-g_0(W)}{1-g_{n}(W)} - 1\right)
        \left(\frac{\bar{Q}_0(0,W)}{\bar{Q}_{n}(0,W)} - 1\right)\right]\right\rvert
        + o_{P}(n^{-1/2})\\
      & \leq \mathbb{E}_{P_0}\left[\frac{W_{j}^2}{\bar{Q}_{n}(1,W)^{2}}
        \left(\frac{g_0(W)}{g(W)} - 1\right)^2\right]^{1/2}
        \mathbb{E}_{P_0}\left[\left(\bar{Q}_0(1,W)
        - \bar{Q}_{n}(1,W)\right)^{2}\right]^{1/2} \\
      & \qquad\qquad + \mathbb{E}_{P_0}\left[\frac{W_{j}^2}{\bar{Q}_{n}(0,W)^{2}}
        \left(\frac{1-g_0(W)}{1-g_{n}(W)} - 1\right)^2\right]^{1/2}
        \mathbb{E}_{P_0}\left[\left(\bar{Q}_0(0,W)
        - \bar{Q}_{n}(0,W)\right)^{2}\right]^{1/2} + o_P(n^{-1/2})\\
      & \overset{\text{a.s.}}{\leq} M \; \mathbb{E}_{P_0}\left[
        \left(\frac{g_0(W)}{g_{n}(W)} - 1\right)^2\right]^{1/2}
        \mathbb{E}_{P_0}\left[\left(\bar{Q}_0(1,W)
        - \bar{Q}_{n}(1,W)\right)^{2}\right]^{1/2} \\
      & \qquad\qquad  + M \; \mathbb{E}_{P_0}\left[
        \left(\frac{1-g_0(W)}{1-g_{n}(W)} - 1\right)^2\right]^{1/2}
        \mathbb{E}_{P_0}\left[\left(\bar{Q}_0(0,W)
        - \bar{Q}_{n}(0,W)\right)^{2}\right]^{1/2} + o_P(n^{-1/2})\\
    \end{split}
  \end{equation*}

  The second equality follows from
  A\ref{ass:rate-consistency-cond-outcome} and the Maclaurin series of
  $\log (x + 1)$. The final inequality follows from
  A\ref{ass:bounded-confounders} and that $Y$ is a positive random
  variable such that $W_{j}^{2} / \bar{Q}(A,W)^{2} \leq M$ almost surely (a.s.).
  The reported result follows by applying the central limit theorem to the first
  term of the von Mises expansion.

\end{proof}

\subsubsection*{Theorem~\ref{thm:id-tte-abs-vip}}
\begin{proof}
  $S_{P_{X,0}}(t|A,W) \equiv S_{0}(t|A,W)$ is immediate from
  A\ref{ass:positivity}, A\ref{ass:no-unmeasured-confounding-t-c}
  and A\ref{ass:censoring-positivity}. Then
  $\Psi^{F}(P_{X,0}; t) = \Psi(P_{0}; t)$.
\end{proof}

\subsection*{Proposition~\ref{prop:eif-tte-abs-vip}}
\begin{proof}
  Using previous results from \citet{moore2011} and the functional delta
  method, we obtain:
  \begin{equation*}
    \begin{split}
      D_{j}(O,P;t)
      & = \frac{d}{d\epsilon} \Psi_{j}(P_{\epsilon}; t) \big|_{\epsilon = 0} \\
      & = \frac{W_{j}}{\mathbb{E}_{P}\left[W_{j}^{2}\right]}
        \left(\int_{0}^{t} d(O, P; u, 1) - d(O, P; u, 0) \; du
        - \Psi_{j}(P; t)W_{j}
        \right) \; .
    \end{split}
  \end{equation*}
\end{proof}

\subsection*{Proposition~\ref{prop:dr-tte-abs-vip-est}}
\begin{proof}
  From the definition of $D_{j}(O,P;t)$ in
  Equation~\eqref{eq:eif-tte-abs-vip}, we find that
  \begin{equation*}
      \mathbb{E}_{P_{0}}\left[D_{j}(O,P;t)\right]
      \propto \mathbb{E}_{P_{0}}\left\{W_{j}\left(
        \int_{0}^{t} (d(O,P;u,1) - d(O,P_{0};u, 1)) -
          (d(O,P;u,0) - d(O,P_{0};u,0)) \; du
        \right)\right\} \;.
  \end{equation*}
\end{proof}
Conditioning on $W$, it suffices to show that
$\mathbb{E}_{P_{0}}[d_{a}(O,P;t) - d_{a}(O,P_{0};t)] = 0$. It follows from
previous results of \citet{vdl2003unified}, \citet{tsiatis2006} and
\citet{cui2022} that this is achieved when
A\ref{ass:cond-surv-consistency} or A\ref{ass:g-consistency} are
satisfied.

\subsection*{Theorem~\ref{thm:asymp-tte-abs-risks-ests}}
\begin{proof}
  The proof is analogous Theorem~\ref{thm:asymp-cont-abs-risk-ests}'s.
  From the results of \citet{moore2011} and the the functional delta method, the
  entropy condition of A\ref{ass:donsker} implies that the third term of
  the von Mises expansion for any given $t$ is $o_{P}(1)$. Then,
  \begin{equation*}
    \begin{split}
      - R(P,P_{0})
      & = \frac{1}{\mathbb{E}_{P_{0}}\left[W_{j}^{2}\right]}
      \mathbb{E}_{P_{0}}\left\{W_{j}\left(
        \int_{0}^{t} (d_{1}(O,P;u) - d_{1}(O,P_{0};u)) -
          (d_{0}(O,P;u) - d_{0}(O,P_{0};u)) \; du
        \right)\right\} \\
      & \overset{\text{a.s.}}{\leq}
        \frac{C}{\mathbb{E}_{P_{0}}\left[W_{j}^{2}\right]}
        \left\lvert \mathbb{E}_{P_{0}}\left\{
        \int_{0}^{t} (d_{1}(O,P;u) - d_{1}(O,P_{0};u)) -
          (d_{0}(O,P;u) - d_{0}(O,P_{0};u)) \; du
        \right\}\right\rvert \\
    \end{split}
  \end{equation*}
  Similar to the proof of Proposition~\ref{prop:dr-tte-abs-vip-est}, it
  suffices to show that the integrand is bounded by $o_{P}(n^{-1/2})$. Indeed,
  this has previously been established under conditions
  A\ref{ass:donsker}, A\ref{ass:non-zero-variance},
  A\ref{ass:bounded-confounders},
  A\ref{ass:consistency-treatment-survival} and
  A\ref{ass:consistency-censoring}. See, for example,
  \citet{vdl2003unified}, \citep{tsiatis2006} and \citet{cui2022}.
\end{proof}

\subsection*{Corollary~\ref{cor:id-tte-rel-vip}}
\begin{proof}
  It follows from the conditions of Theorem~\ref{thm:id-tte-abs-vip} that
  $\Gamma^{F}(P_{X,0})$ is equal to $\Gamma(P_{0})$.
\end{proof}

\subsection*{Proposition~\ref{prop:eif-tte-rel-vip}}
\begin{proof}
  Again relying on the point mass contamination approach, we obtain:
  \begin{equation*}
    \begin{split}
      D_{j}(O,P;t)
      & = \frac{d}{d\epsilon} \Gamma_{j}(P_{\epsilon}; t) \big|_{\epsilon = 0} \\
      & = \frac{W_{j}}{\mathbb{E}_{P}\left[W_{j}^{2}\right]}
        \left(
        \frac{2A-1}{Ag(W) + (1-A)(1-g(W))}
        \int_{0}^{t}
        \frac{I(\tilde{T} \geq u)}
        {c(u_{-}|A,W) S(u|A,W)}
        \left(I(T=u)-\lambda(u|A,W) \; du\right) \right.\\
      & \qquad\qquad\qquad\qquad + \log \frac{S(t|1,W)}{S(t|0,W)}
        - \Gamma(P;t)W_{j}\Bigg) \;.
    \end{split}
  \end{equation*}
\end{proof}

\subsection*{Proposition~\ref{prop:dr-tte-rel-vip-est}}
\begin{proof}
  By the definition of Equation~\eqref{eq:eif-tte-rel-vip}, we have that
  \begin{equation*}
    \begin{split}
      \mathbb{E}_{P_{0}}\left[D_{j}(O,P;t)\right]
      & \propto \mathbb{E}_{P_{0}}\left\{W_{j}\left(
        \left(\frac{g_{0}(W)}{g(W)} - \frac{1-g_{0}(W)}{1-g(W)}\right)
        \int_{0}^{t}
        \frac{c_{0}(u_{-}|A,W)}{c(u_{-}|A,W) S(u|A,W)}
        \left(\lambda_{0}(u|A,W)-\lambda(u|A,W)\right) \; du \right.\right.\\
      & \qquad\qquad\qquad\qquad\qquad \left.\left.
        + \log\frac{S(t|1,W)}{S(t|0,W)}
        - \log\frac{S_{0}(t|1,W)}{S_{0}(t|0,W)}\right)\right\} \\
    \end{split}
  \end{equation*}
  Then $\mathbb{E}_{P_{0}}\left[D_{j}(O,P;t)\right]=0$ under
  A\ref{ass:cond-surv-consistency}.
\end{proof}

\subsection*{Theorem~\ref{thm:asymp-tte-rel-risks-ests}}
\begin{proof}
  Again, the entropy condition of A\ref{ass:donsker} implies that the third term
  of the von Mises expansion is $o_{P}(1)$. Studying the remainder term, we find
  that
  \begin{equation*}
    \begin{split}
      P- R(P_0,\hat{P}_{n})
      & = \frac{1}{\mathbb{E}_{P_{0}}\left[W_{j}^{2}\right]}
        \mathbb{E}_{P_{0}}\left\{W_{j}\left(
        \frac{2A-1}{Ag_{n}(W) + (1-A)(1-g_{n}(W))} \right.\right.\\
      & \qquad\qquad\qquad\qquad \left.\left.
        \int_{0}^{t}
        \frac{I(\tilde{T} \geq u)}
        {c_{n}(u_{-}|A,W) S_{n}(u|A,W)}
        \left(I(T=u)-\lambda_{n}(u|A,W)\right) \; du \right.\right.\\
      & \qquad\qquad\qquad\qquad\qquad \left.\left.
        + \log\frac{S_{n}(t|1,W)}{S_{n}(t|0,W)}
        - \log\frac{S_{0}(t|1,W)}{S_{0}(t|0,W)}\right)\right\} \\
      & = \frac{1}{\mathbb{E}_{P_{0}}\left[W_{j}^{2}\right]}
        \mathbb{E}_{P_{0}}\left\{W_{j}\left(
        \left(\frac{g_{0}(W)}{g_{n}(W)} - \frac{1-g_{0}(W)}{1-g_{n}(W)}\right)
        \right.\right. \\
      & \qquad\qquad\qquad\qquad \left.\left.
        \int_{0}^{t}
        \frac{c_{0}(u_{-}|A,W)}{c_{n}(u_{-}|A,W) S_{n}(u|A,W)}
        \left(\lambda_{0}(u|A,W)-\lambda_{n}(u|A,W)\right) \; du \right.\right.\\
      & \qquad\qquad\qquad\qquad\qquad \left.\left.
        + \log\frac{S_{n}(t|1,W)}{S_{n}(t|0,W)}
        - \log\frac{S_{0}(t|1,W)}{S_{0}(t|0,W)}\right)\right\} \\
      & = \frac{1}{\mathbb{E}_{P_{0}}\left[W_{j}^{2}\right]}
        \mathbb{E}_{P_{0}}\left\{W_{j}\left(
        \left(\frac{g_{0}(W)}{g_{n}(W)} - \frac{1-g_{0}(W)}{1-g_{n}(W)}\right)
        \right.\right. \\
      & \qquad\qquad\qquad\qquad \left.\left.
        \int_{0}^{t}
        \frac{c_{0}(u_{-}|A,W)}{c_{n}(u_{-}|A,W) S_{n}(u|A,W)}
        \left(\lambda_{0}(u|A,W)-\lambda_{n}(u|A,W)\right) \; du \right.\right.\\
      & \qquad\qquad\qquad\qquad\qquad \left.\left.
        + \left(\frac{S_{n}(t|1,W)}{S_{0}(t|1,W)} - 1\right)
        - \left(\frac{S_{0}(t|0,W)}{S_{n}(t|0,W)}-1\right)\right)\right\}
        + o_{P}(n^{-1/2})\\
      & = \frac{1}{\mathbb{E}_{P_{0}}\left[W_{j}^{2}\right]}
        \mathbb{E}_{P_{0}}\left\{W_{j}\left(
        \left(\frac{g_{0}(W)}{g_{n}(W)}\frac{c_{0}(t|A,W)}{c_{n}(t|A,W)} - 1\right)
        \left(\frac{S_{0}(t|1,W)}{S_{n}(t|1,W)} - 1\right) \right.\right. \\
      & \qquad\qquad\qquad\qquad\qquad \left.\left.
        + \left(\frac{1-g_{0}(W)}{1-g_{n}(W)}\frac{c_{0}(t|A,W)}{c_{n}(t|A,W)}-1\right)
        \left(\frac{S_{0}(t|A=0,W)}{S_{n}(t|A=0,W)} - 1\right)
        \right)\right\} + o_{P}(n^{-1/2})\\
      & \overset{\text{a.s.}}{\leq}
        \frac{C}{\mathbb{E}_{P_{0}}\left[W_{j}^{2}\right]}
        \Bigg\{
        \mathbb{E}_{P_{0}}\left[
        \left(\frac{g_{0}(W)c_{0}(t|A,W)-g_{n}(W)c_{n}(t|A,W)}{g_{n}(W)c_{n}(t|A,W)}\right)^{2}
        \right]^{1/2} \\
      & \qquad\qquad\qquad\qquad\qquad\qquad
        \mathbb{E}_{P_{0}}\left[
        \left(S_{0}(t|1,W) - S_{n}(t|1,W)\right)^{2}
        \right]^{1/2} \\
      & \qquad\qquad\qquad\qquad
        + \mathbb{E}_{P_{0}}\left[
        \left(\frac{(1-g_{0}(W))c_{0}(t|A,W)-(1-g_{n}(W))c_{n}(t|A,W)}
        {(1-g_{n}(W))c_{n}(t|A,W)}\right)^{2}
        \right]^{1/2} \\
      & \qquad\qquad\qquad\qquad\qquad\qquad
        \mathbb{E}_{P_{0}}\left[\left(S_{0}(t|0,W) - S_{n}(t|0,W)\right)^{2}
        \right]^{1/2}\Bigg\}  + o_{P}(n^{-1/2}) \; .\\
    \end{split}
  \end{equation*}
  The stated result follows.

\end{proof}

\FloatBarrier

\bibliographystyle{unsrtnat}
\bibliography{references}

\end{document}